\documentclass[journal=aamick, manuscript=article, layout=twocolumn]{achemso}
\usepackage[version=3]{mhchem}
\usepackage[utf8]{inputenc}
\usepackage[T1]{fontenc}
\usepackage{lipsum}
\usepackage{xcolor}
\usepackage{makecell}

\usepackage{pdfpages}

\setkeys{acs}{maxauthors=0}

\let\oldmaketitle\maketitle
\let\maketitle\relax
\author{Carlos Mera Acosta}
\email{carlos.acosta@ufabc.edu.br}
\affiliation[UFABC]{Federal University of ABC (UFABC), 09210-580, Santo Andr\'e, S\~ao Paulo, Brazil}
\altaffiliation{These authors contributed equally to this work.}
\author{Elton Ogoshi}
\email{elton.ogoshi@ufabc.edu.br}
\affiliation[UFABC]{Federal University of ABC (UFABC), 09210-580, Santo Andr\'e, S\~ao Paulo, Brazil}
\altaffiliation{These authors contributed equally to this work.}
\author{Jose Antonio Souza}
\affiliation[UFABC]{Federal University of ABC (UFABC), 09210-580, Santo Andr\'e, S\~ao Paulo, Brazil}
\author{Gustavo M. Dalpian}
\email{gustavo.dalpian@ufabc.edu.br}
\affiliation[UFABC]{Federal University of ABC (UFABC), 09210-580, Santo Andr\'e, S\~ao Paulo, Brazil}

\title[Machine learning of 2D magnetic materials]
{Machine Learning Study of the Magnetic Ordering in 2D Materials }
\abbreviations{ML, 2D, TMDs, DFT, PEC}

\begin{document}
%
\twocolumn[
\begin{@twocolumnfalse}
\oldmaketitle
\begin{abstract}
Magnetic materials have been applied in a large variety of technologies, from data storage to quantum devices. The development of 2D materials has opened new arenas for magnetic compounds, even when classical theories discourage their examination.
Here we propose a machine-learning-based strategy to predict and understand magnetic ordering in 2D materials. This strategy couples the prediction of the existence of magnetism in 2D materials using random forest and the SHAP method with material maps defined by atomic features predicting the magnetic ordering (ferromagnetic or antiferromagnetic). While the random forest model predicts magnetism with an accuracy of 86\%, the material maps obtained by the SISSO method have an accuracy of about 90\% in predicting the magnetic ordering. Our model indicates that 3$d$ transition metals, halides, and structural clusters with regular transition metals sublattices have a positive contribution in the total weight deciding the existence of magnetism in 2D compounds. This behavior is associated with the competition between crystal field and exchange splitting. The machine learning model also indicates that the atomic SOC is a determinant feature for the identification of the patterns separating ferro- from antiferromagnetic order. 
The proposed strategy is used to identify novel 2D magnetic compounds which, together with the fundamental trends in the chemical and structural space, paves novel routes for experimental exploration. 
\vspace{-0.5cm}
\end{abstract}
\end{@twocolumnfalse}
]
\section{Introduction}
The first synthesis of graphene opened a new era for two-dimensional (2D) materials~\cite{Novoselov2004}. These materials, with strong in-plane and weak interlayer interactions, enclose nearly all electronic and optical phenomena found in solids. The subsequent experimental realization and theoretical insights of other 2D materials~\cite{Splendiani2010} have influenced and transformed our fundamental understanding of materials properties. It also envisaged and put forward several technological applications in many areas of condensed matter physics, chemistry, and materials science.
In this 2D materials revolution, some breakthroughs were achieved including the discovery of superconductivity in rotated bilayer graphene~\cite{cao2018}, the theoretical description of symmetry protected topological phases in bismutene \cite{kou2017} and the extremely high electron mobilities in phosphorene\cite{Das2014}.
Although 2D materials physics overlaps many other well-established areas, historically, it has been anticipated the existence of mutual exclusion with magnetism, i.e., long-range ordering of spin and orbital magnetic moments~\cite{Mermin1966}. This brings limitations for the integration of 2D materials in magnetic devices. 

The study of magnetism and magnetic materials goes back to the beginning of the theoretical foundations of physics and chemistry. However, even though there have been theoretical predictions of magnetic stable monolayers\cite{magnet_book}, a definitive experimental observation of magnetic order at finite temperature in 2D layered materials was only reported  recently\cite{Mak2019,Ningrum2020,Vedmedenko2020,Sethulakshmi2019}. Specifically, scanning magneto-optic Kerr microscopy measurements revealed intrinsic long-range ferromagnetic order in CrGeTe$_3$ Van der Waals (VdW) atomic layers\cite{Gong2017}. Similarly, Bevin Huang et al., demonstrated for the first time the existence of ferromagnetism in CrI$_3$ VdW crystals down to the monolayer limit\cite{Huang2017}. The discovery of 2D magnetism has not only opened a new path for potential revolutionary spintronic~\cite{Feng2017} and valleytronic~\cite{Farooq2019} applications but also raised fundamental questions about the understanding and prediction of ferromagnetic (FM) and antiferromagnetic (AFM) ordering in 2D compounds.

In three-dimensional bulk materials, the nature of magnetic ordering ground state is determined by collective mechanisms involving magneto-crystalline anisotropy arising from the spin-orbit coupling and how magnetic moments (S and L) are effectively coupled by exchange interactions. Furthermore, the interplay and coexistence of exchange interactions can give rise to not only FM and AFM orderings, but also to a very rich variety of different magnetic states even in compounds with similar crystal structures.\cite{dalpian2006} 
In the simplest scenario, two identical transition metal ions with unpaired electrons interact through the isotropic Heisenberg Hamiltonian $\hat{H}=-JS_{1}S_{2}$, where the electron spins $S_{1}$ and $S_{2}$ are coupled by the exchange interaction parameter $J$. 
In this isotropic magnet, depending on the extent of delocalization of the magnetic moments, the magnetic order can be described by four exchange interactions – indirect, itinerant, direct, and super exchanges\cite{magnet_book}. These exchange interactions can coexist in a given compound, suggesting that even without magneto-crystalline anisotropy, the magnetic ordering is determined by competing effects. Indeed, the complexity in predicting magnetic ordering has been recently recalled: "\textit{the prediction of the magnetic state based solely on chemical and structural information is a delicate exercise}"\cite{Nelson2019}. The description of magnetic ordering in 2D materials is even more complex. Specifically, in 1966, based on the isotropic Heisenberg model, Mermin and Wagner demonstrated that, unlike bulk compounds, in 2D materials long-range magnetic order is suppressed by thermal fluctuations. This suggests that 2D materials can only exhibit magnetic order in the presence of large magneto-crystalline anisotropy and hence, the magnetism in 2D and bulk materials are supposed to have different physical mechanisms.\cite{Mermin1966}    
Indeed, in general, there is no rule, as a knowledge base, that \textit{a priori} determines the magnetic behavior in 2D materials or a trend of AFM or FM magnetic orderings in the feature space of atomic properties. For instance, 2D magnetic semiconductors can violate the Goodenough-Kanamori rules for superexchange interactions\cite{Kabiraj2020} that discriminate FM from AFM orders according to the angles between $d$-orbitals from TM and $p$-orbitals from oxygen anions\cite{Goodenough1958,Kanamori1960}. Clear and well-defined patterns would facilitate the search for stable AFM and FM 2D materials. In the ideal scenario, one wishes to determine a set of stable AFM and FM configurations for a given combination of atoms with a specific stoichiometry and crystal symmetry without additional calculations. Our work thus aims to solve this problem using machine learning algorithms.

In recent years, much effort has been invested in the systematic prediction of magnetic two-dimensional systems~\cite{vanGog2019,ZhuYu2018,Kumar2017,JiangZhou2018,ZhuangHoulong2015,MiaoNaihua2018,Gonzalez_2019,KanM2013,GengJiazhong2020}, coming mainly from density functional theory (DFT) calculations. These predictions are based on a \textit{direct approach} that involves the calculation of all possible candidates. This trial-and-error process can be time-consuming, and expensive. 
Daniele Torelli et al., used the computational 2D materials database (C2DB)~~\cite{Haastrup2018,gjerding2021recent} to search for new ferromagnetic 2D materials based on the spinwave gap as a descriptor that accounts for the role of magnetic anisotropy, finding 17 novel insulating materials that exhibit magnetic order at finite temperatures~\cite{Torelli_2019}. Similarly, a computational screening for 2D magnetic materials based on experimental bulk compounds that are exfoliable into 2D derivatives~\cite{Mounet2018} found 85 ferromagnetic and 61 antiferromagnetic materials for which DFT calculated magnetic exchange and anisotropy parameters are reported~\cite{Torelli2020}. Finally, Arnab Kabiraj et al., developed a fully automated code to perform DFT calculations followed by Heisenberg model-based Monte Carlo simulations to estimate the Curie temperature from the crystal structure, predicting 26 materials with a Curie point beyond 400 K~\cite{Kabiraj2020}.

Here we propose a data-driven strategy to explore the magnetism in 2D materials: contrary to the commonly employed \textit{direct approach}, we aim to explore the material-to-attribute connection to provide the simplest correlation between magnetic ordering and features, e.g., composition, crystal symmetry, and atomic properties\cite{Schleder2019r,Dalpian2019,acosta2018}. Specifically, we train machine learning algorithms using a recently created database~\cite{Kabiraj2020} of 2D magnetic materials to obtain descriptors that are capable of classifying materials as non-magnetic, ferromagnetic, or antiferromagnetic. Our strategy is divided into two main steps, namely: i) we first develop a random forest model to separate magnetic compounds from non-magnetic ones based on trends in the crystal structure and atomic composition; and ii) based on the sure independence screening and sparsifying operator (SISSO) method~\cite{Ouyang2018}, we find a mathematical model (i.e., a function of the atomic features) that uses the composition to provide a materials map with defined regions for AFM and FM 2D materials. The accuracy in the classification of 2D AFM and FM is about 90\% in the training set. We find that the magnetic ordering is decided by features involving the lowest-
occupied Kohn-Sham eigenvalues for the
cations and anions that constitute the target material, their electron affinity, and atomic radii of $p$-orbitals. However, different rules involving these features are observed for materials that belong to different space groups, clearly indicating the complexity of this problem.

\section{Machine learning models, results and discussion}

\begin{figure*}[h]
    \centering
    \includegraphics[width=\linewidth]{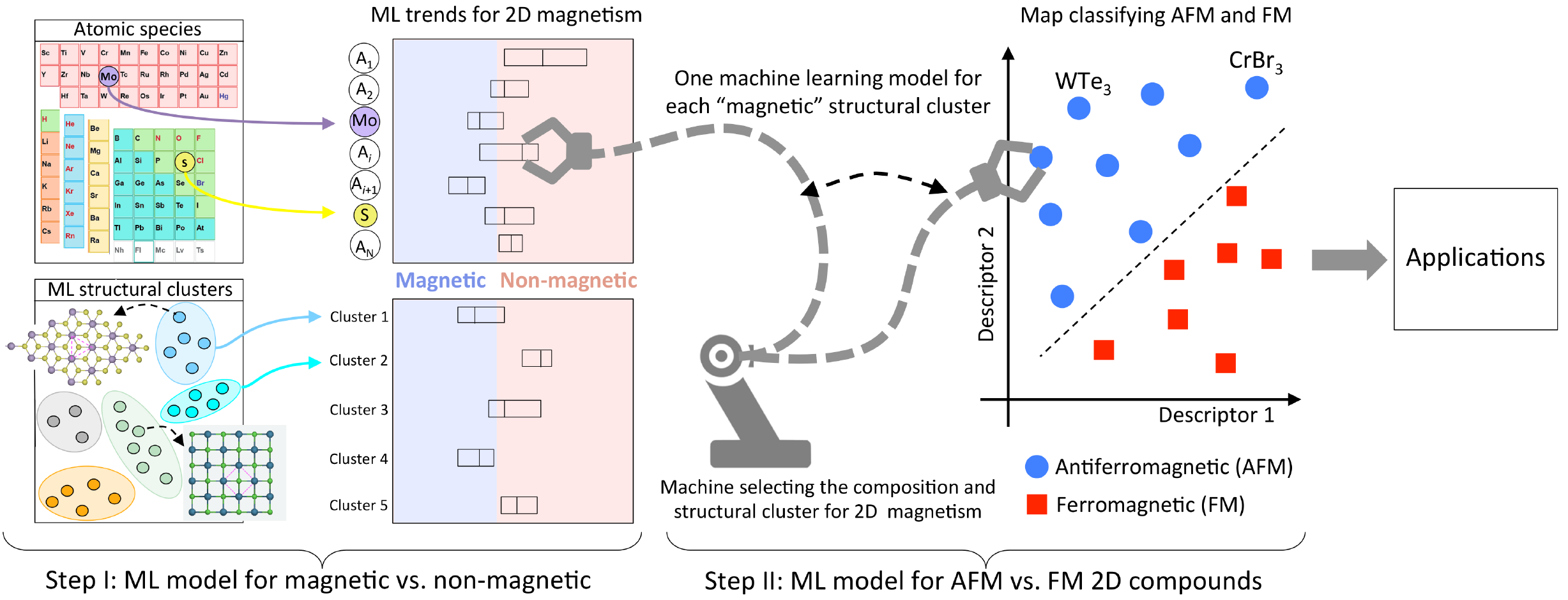}
    \caption{Illustration of the proposed machine learning strategy to determine and predict the existence of magnetism (step I) in a given 2D compound and its specific magnetic ordering (step II). The starting point for this strategy is the clustering of atomic species and structures. Atomic species are intrinsically clustered according to the organization of the periodic table, while an ML model is performed to group structures according to the possible composition, crystal point group, and local distortions. This cluster is then used to identify the tendency of a specific element and structure to form a magnetic (blue) or non-magnetic (red) 2D structure. If the specific combination of atoms and structures for the 2D compound tends to be magnetic, we identify its position in a materials map separating antiferromagnetic from ferromagnetic ordering. These positions are computed by means of atomic features of the constituent atoms.  Finally, this ML strategy allows us to predict and select compounds for specific applications in electronics and spintronics.}
    \label{fig1}
\end{figure*}

In the ideal scenario, one would like to have a model determining the existence and type of magnetic ordering in an arbitrary combination of Atoms, in a specific Composition, and Symmetry (ACS). In this section, we present the proposed strategy to solve the two above-noted problems. First, as presented in Figure\ref{fig1}, the proposed scheme divides the problem into two steps, namely: (I) classification for magnetic vs non-magnetic materials and  (II) classification for AFM vs FM ordering in 2D compounds. In order to input the ACS attributes into the ML model, we start by encoding the materials' compositions and crystal structure information into proper machine-readable features. In this first step, these features are used to study the distribution of magnetism on the space of different two-dimensional materials.
Compounds are then classified as magnetic (M) or non-magnetic (NM) according to the DFT-calculated magnetic ground state. The second step is the machine learning classification that separates AFM from FM 2D materials to construct a materials map of compounds with magnetic ordering. The axis of this materials map (see Figure~\ref{fig1}) are given by descriptors that are functions of the atomic properties $D_{n}=f(\xi_1,\xi_2,..., \xi_n)$, where $n$ is the descriptor dimension and $\{\xi_m\}$ are the $m$ atomic properties identifying the compound. As illustrated in Figure~\ref{fig1}, the prediction of novel magnetic 2D compounds starts by selecting both the atomic elements and structural clusters with a larger tendency to be magnetic. Then, based on the selected atoms and crystal point group, the mathematical descriptors $D_{n}$ are calculated for all possible compositions formed by the selected atoms in the selected structure; the position in the materials map will indicate if the 2D materials candidates are either AFM or FM.  
In ML models for materials, three components are needed: available data, a numerical representation of materials, and learning algorithms and their optimization~\cite{Schleder2019r}.



The starting point for both ML models described in Figure~\ref{fig1} is the C2DB~\cite{Haastrup2018} database of bidimensional materials, which contains a total of 3814 entries in its 2020 version. The C2DB database was built with DFT calculations of experimentally known 2D structural prototypes combinatorially decorated with different ions from the periodic table. Each entry in the C2DB database is a combination of a crystal structure, a composition, and a magnetic order. We filtered out materials with heat of formation~>~-25 meV/atom, so the dataset contains only materials that are thermodynamically stable. Additionally, we decided to exclude the materials which composition did not include a transition metal. This series of filters resulted in a dataset with 2205 entries. By grouping the entries by crystal structure and composition, we ended with a total of 1845 materials, each with (possibly) different magnetic orderings.



\subsection*{Random forest classification of magnetic and non-magnetic 2D compounds}

 The random forest classification scheme of magnetic (M) and non-magnetic(NM) compounds is discussed below:

{\it Available magnetic and non-magnetic 2D compounds entries}: To extract trends differentiating M and NM, we propose the training of an interpretable ML algorithm using the 850 M and 2569 NM compounds from the C2DB~\cite{Haastrup2018}.
The DFT calculated total energy provided by the C2DB for NM ($E_{FM}$) and M ($E_{NM}$) phases allows to identify the ground state for the initial set of 1845 2D compounds. 
ML models and general trends for the NM and M behavior in the chemical and structural space requires to eliminate ambiguous entries, i.e., those with a relatively small energy difference between different magnetic configurations, $\Delta E_{\text{M}-\text{NM}}=E_{\text{M}}-E_{\text{NM}}$. We thus exclude compounds with $|\Delta E_{\text{M}-\text{NM}}|< 0.01$ eV/atom, resulting in a dataset with 1713 compounds, with 474 M and 1239 NM.


{\it Numerical representation of magnetic and non-magnetic compounds}: 
The descriptors used to characterize the chemical space were simple boolean features that are set to True (False) when the element is (not) present in the composition. 
To get to a structural descriptor, we used a non-supervised approach to cluster the crystal structures, and then we used these cluster labels as a categorical feature.   
The C2DB actually provides a label for grouping different structures and it has a format that uses the material's stoichiometry, crystal space~group, and the set of occupied Wyckoff positions. The H-phase of \ce{MoS2} is labeled as \ce{AB2}-187-ai, for example. Even though this label is very specific, unfortunately we observed that it was not qualitatively univocal, i.e. in some cases it grouped different structures into the same label. In order to bypass that, we used CrystalNN crystal fingerprint~\cite{crystalnn} to find an embedding that deeply characterizes the structural differences of all materials. The CrystalNN is based on the evaluation of every site of the crystal regarding different local geometrical environments (e.g octahedral, tetrahedral). The outcome is then a 244-dimensional vector for each material. Using these vectors in an embedding method such as the T-SNE~\cite{tsne}, and then automatically clustering this embedding using DBSCAN~\cite{dbscan}, we were able to get to a set of 51 structural groups. The Supporting Information provides some further details on this process. For this specific analysis, this kind of feature has shown to be more efficient than the labeling from C2DB or simply grouping by space group or point group.

{ \it Learning algorithm and its optimization: Random forest classifier.} With the huge variety of models one can use to make classifications\cite{Schleder2019}, we chose the non-linear Random Forest (RF) model due to its simplicity and ease of interpretation. Another great advantage of RFs is that, by using the out-of-bag methodology (OOB)~\cite{oob}, it is possible to use 100\% of the data for both training and for validating the model. Details regarding its implementation and hyperparameters are presented in the Supporting Information. Table 1 shows the RF accuracy results for the 1684 materials.
As can be observed, the out-of-bag F1 score is 86\%, which indicates that by using very simple compositional and structural features one is able to delineate bidimensional materials which have magnetic order with good accuracy.

\begin{table}[]
\setlength\tabcolsep{9pt} 
\small
\centering
\caption{Random Forest Out-of-bag (OOB) validation results. The F1 score is the best performance metric as the dataset has two unbalanced classes, with M as the minority class. M/NM Accuracy stands for the RF individual class accuracy, i.e. how many materials of these classes are correctly classified by the model.}
\label{tab:rf_results}
\begin{tabular}{lcc}
\multicolumn{1}{c}{} & \multicolumn{1}{r}{}\\ \hline
Number of materials & 1713\\
M Proportion        & 25.8\%\\
Accuracy            & 92.5\%\\
F1 score            & 86.0\%\\
M Accuracy          & 84.5\%\\
NM Accuracy         & 95.9\%\\ \hline
\end{tabular}
\end{table}

\begin{figure*}[h!]
    \centering
    \includegraphics[width=15cm]{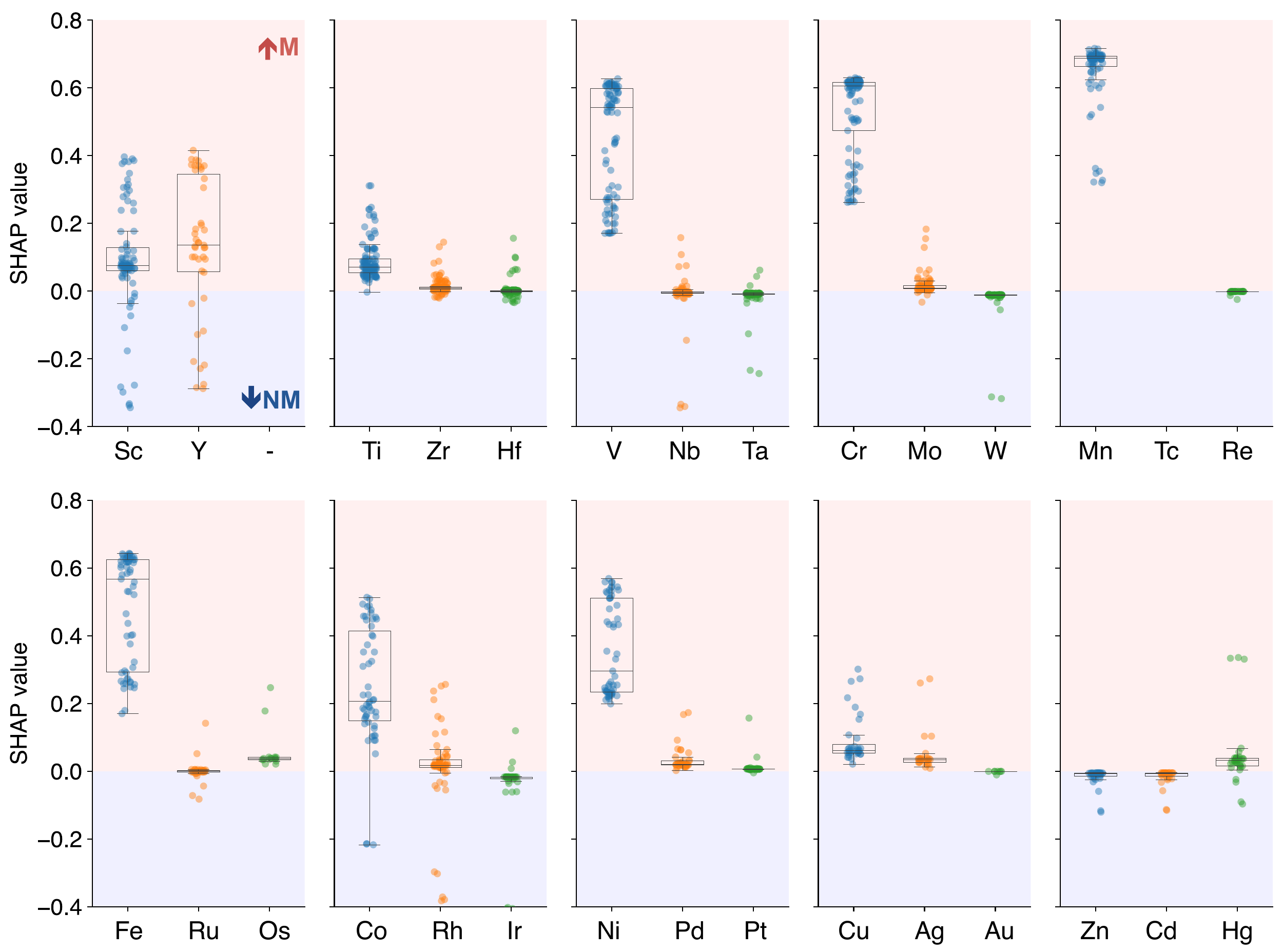}
    \caption{Strip plots of SHAP values of compositional features for all TMs. Each individual plot represents a group, with color encoding the period in the periodic table. Each dot represents a calculated SHAP value for an instance/material which contains the TM in its composition. Positive (Negative) SHAP values, indicated by red (blue) background, mean that the presence of the specific TM contributes to the model to classify the material as M (NM). SHAP values concentrated around zero indicates that having the TM in the composition does not affect the model classification.}
    \label{fig:TM_shap}
\end{figure*}

{\it Trends for M vs. NM behavior in the space of chemical species}: The next step was to rationalize what the RF models have learned in order to extract trends. For this task, we used the popular SHAP~\cite{shap} methodology used to explain tree-based ML models. This method is based on using game theory's Shapley values~\cite{roth1988shapley} (or SHAP values, in this context) for assigning the impact of features on the model's predictions at the instance level. These values are calculated for each feature value and for each specific instance in the dataset: by using all possible combinations of the set of features, i.e. the feature power set containing $2^n$ different feature combinations ($n$: number of features), the method consists in fitting separated one-instance models to the RF model output, making a total $2^n$ fittings for each instance. From this set of fittings, it is possible to calculate the feature Shapley values, which evaluate numerically how the feature contributes to the RF output.

In the context of an NM vs M binary classification, SHAP values can be negative or positive if the feature locally contributes to a classification of NM or M, respectively. Their absolute values are directly correlated with how important the feature value was to the model output. Thus one is able to investigate the general influence of a feature on the model by analyzing the distribution of all its SHAP values. 

We chose the SHAP methodology over traditional model-level importance features, such as Permutation Feature Importance (PFI), because the latter does not provide class-specific values. While PFI directly provides a single-valued weight for each feature (on which the greater the value, the more important the feature is), it fails on discriminating how a feature helps to classify the material as NM or M. In general, model-level importance features lacks the description of how the individual feature values affect the classification, while SHAP, as an instance-level method, covers this shortcoming by showing how each feature value contributes locally for the model output.

\begin{figure}[h!]
    \centering
    \includegraphics[width=8cm]{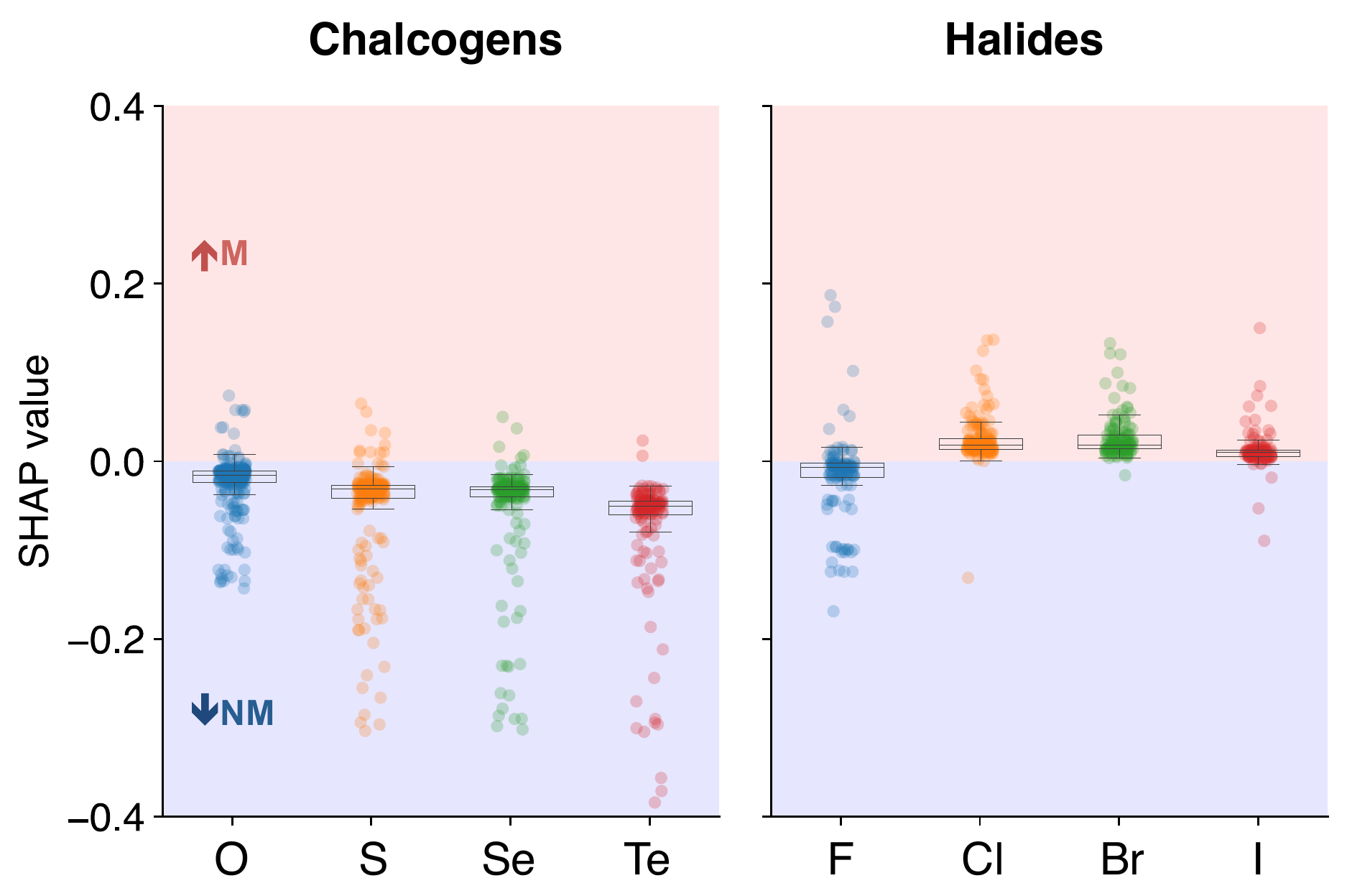}
    \caption{Strip plots of SHAP values of compositional features for anions (chalcogens and halogens).}
    \label{fig:anions_shap}
\end{figure}

Figures~\ref{fig:TM_shap} and~\ref{fig:anions_shap} show the distribution of compositional features' SHAP values. As the C and N groups are not statically representative, and for the sake of simplicity, we present only the plots for transition metals (TMs) and for anion groups (chalcogens and halogens). The Supporting Information provides the full plots for the cited remaining periodic table groups (Figure S2).

From Figure~\ref{fig:TM_shap} we can see a clear trend regarding the effect of the period of the TM. While 4d (orange) and 5d (green) TMs show neutral contributions to the model, visually denoted by their SHAP values distributions concentrated around 0.0, the 3d (blue) TMs show very positive SHAP values, with the highest values happening for \ce{Mn}, \ce{Fe}, \ce{Cr}, \ce{V}, \ce{Ni} and \ce{Co}. This means that the presence of these
elements had a huge impact on the model classification of the compounds into magnetic (M). 

The difference between 3d TM and 4d, 5d TMs also happens in tridimensional materials and can be explained by the crystal/ligand field splitting of d-levels. 3d TMs usually have a smaller crystal splitting of d-levels, which are associated with FM or AFM magnetic order, as we discuss in the next section.


The influence of different anions (Figure~\ref{fig:anions_shap}) in classifying a material into M or NM was also evaluated. We see that compared to the TMs, the anions SHAP values distributions are concentrated around much lower absolute values, indicating that they are less determinant in the classification process. However, we can qualitatively see that the model associates the presence of chalcogens with NM order, while the presence of halides is associated with M. This means that the reduced coordination of halides presents a higher correlation with magnetic ordering in two-dimensional materials.

\begin{figure*}[h!]
    \centering
    \includegraphics[width=15cm]{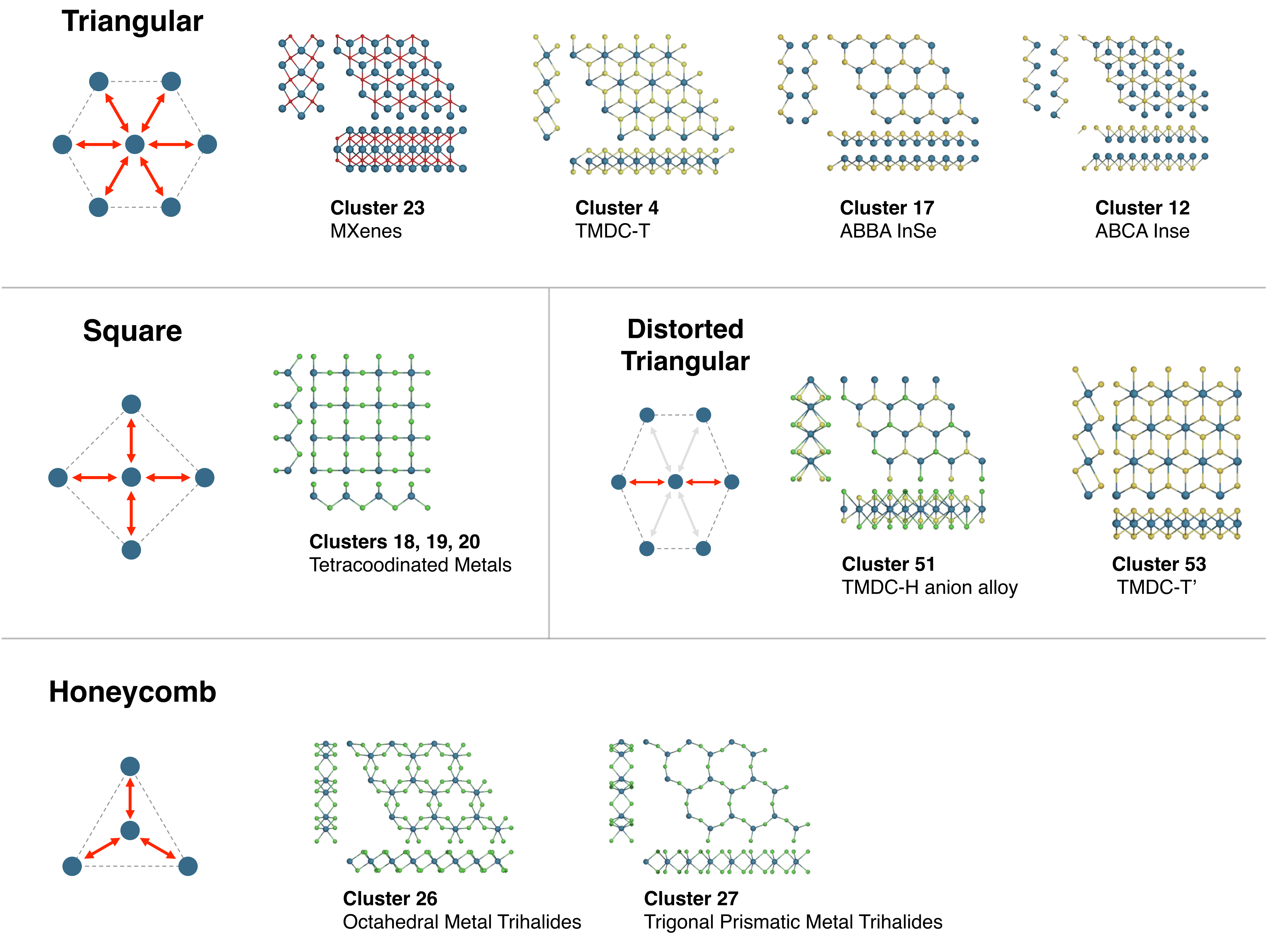}
    \caption{All structural clusters with SHAP values distributions correlated with a magnetic classification. They can be divided into four main groups regarding the transition metal sublattice: Triangular, Square, Honeycomb, and Distorted Triangular.
    }
    \label{fig:clusters}
\end{figure*}

{\it Trends for M vs. NM behavior in the space of structural clusters}: The categorical structural feature has a total of 51 labels corresponding to 51 qualitatively different structural groups. From the SHAP distributions plots for structural clusters, presented in the Supporting Information (Figure S3), we see that some of them present large absolute SHAP values and thus high influence on the RF model prediction. Figure~\ref{fig:clusters} shows the general crystal structure of selected clusters that contributed to an M classification. 

The selected structural clusters can be separated into four main groups regarding their respective TM sublattices: Honeycomb, Square, Triangular, and Distorted triangular. By verifying the recent literature on magnetism in two-dimensional materials~\cite{jiang2021recent}, one can see that most of the compounds where magnetism has been verified (experimentally or computationally) present a crystal structure that fits into the general structures of the clusters shown in Figure~\ref{fig:clusters}.

The first experimental observations of magnetic ordering in the monolayer limit support the structural trend for Honeycomb TM sublattices, for example, with \ce{FePS3} in 2016~\cite{Lee2016} and \ce{CrI3} (cluster 26) in 2017~\cite{Huang2017} presenting AFM and FM order, respectively. Other monolayers with the {\it cluster 26} structure and with composition \ce{MX3} (M = Cr, Mn; X = F, Cl, Br, I) also have been studied and theoretically predicted to present intrinsic ferromagnetism~\cite{zhang2015robust,sun2018prediction}. Calculations of \ce{CoBr3} in the {\it cluster 27} crystal structure also showed that it presents intrinsic FM order~\cite{zhang2019two}.

The Triangular TM sublattice result is also supported by experimental and computational reports, such as the experimental verification of ferromagnetism at room temperature in \ce{VSe2}~\cite{Bonilla2018} and \ce{MnSe2}~\cite{OHara2018}, both in the T-phase of TMDCs (cluster 4); the FM phase prediction by calculations of \ce{ScCl}, \ce{YCl}, \ce{LaCl} in the {\it cluster 12} structure ~\cite{JiangZhou2018,wang2018high}; and the theoretical ground state of AFM for \ce{Ta3C2}~\cite{lane2013correlation} and FM for \ce{Cr3C2}~\cite{zhang2017robust} and \ce{Ti3C2}~\cite{wu2015theoretical} in the pristine MXene crystal structure ({\it cluster 23}).

Magnetic two-dimensional materials with Square or Distorted Triangular sublattices, as far as we know at the moment, remain to be further investigated. Even though the study of specific compounds with these structures was not found, the training data from C2DB suggests that these geometries are viable for magnetic order and for future experimental verification.

With the exception of the Distorted Triangular TM sublattice, all groups present a regular transition metal 2D sublattice; they represent the only three ways one can regularly tile a 2D plane, with honeycomb, triangular or square geometries.

In contrast to this regularity, the only structural cluster which contributed effectively to an NM classification was the one with the general crystal structure of H-TMDC transition metal alloys ({\it cluster 33}). These specific alloys have a stoichiometry of \ce{ABX4} or of \ce{AB3X8}, where \ce{A} and \ce{B} are transition metals from periods 4 and 6, and \ce{X} is a chalcogen. Something particular of this structure is that the alloying with \ce{A} TM breaks the chemical regularity of the \ce{B} TM and, in some cases, it even breaks the structural regularity by breaking the system mirror symmetry, and thus distorting the TM plane.

From these results, the main structural trend suggested is the need for geometric regularity of the TM sublattice for the emergence of intrinsic magnetic ordering. This regularity is possibly correlated with long-range exchange interactions between TM nearest neighbors.


\subsection*{Machine learning classification of AFM and FM 2D compounds}

The tendency in the feature space defined by the atomic properties and crystal symmetries of the magnetic behavior allows us to understand which atomic combinations and structures are prone to present a magnetic ordering, however, it does not indicate what type of order is expected. In this section, we thus focus on the machine learning distinction of magnetic phases such as FM and AFM. We first describe the three components that are needed for machine learning models: available data, a numerical representation of materials, and learning algorithms.

\textit{Available 2D compounds with magnetic ordering:}
The computational 2D materials database (C2DB) \cite{Haastrup2018} is also the source of the data set used to train the machine learning model discriminating AFM from FM ordering. In the C2DB, the three considered magnetic phases (i.e., non-magnetic, ferromagnetic or anti-ferromagnetic) are not always calculated for a given compound. Indeed, there are 3052 compounds for which only one of these magnetic phases is reported: 2533 NM, 446 FM, and 73 AFM. Similarly, NM and FM phases are evaluated for 195 compounds, FM and AFM are calculated for 147 2D materials, and only 14 compounds with reported total energy for the NM and AFM phases. However, for the 875 compounds with non-zero local magnetic moments, the DFT calculations are based on a \textit{one supercell model}, i.e., the considered AFM and FM phases are restricted to the smaller unit cell with at least two transition metals. Calculating all possible symmetry allowed magnetic orderings, Arnab Kabiraj {\it et al}~\cite{Kabiraj2020} found that from the 788 DFT calculated FM phases, 368 are actually AFM. Naturally, it is also possible that some compounds are misclassified as NM, which can explain the obtained accuracy in the magnetic vs non-magnetic classification in the previous section. Previous machine learning classification for magnetic ordering based on the C2DB intrinsically has the bias of this DFT calculations~\cite {Miyazato_2018}. Here, we restrict the training set to the database of 525 magnetic order compounds from Ref.~\citenum{Kabiraj2020}. 
Even though at a first glance this seems to be a relatively small dataset for training, this has shown to be sufficient for other ML analysis in Materials Science \cite{Jena2021}. Also, for the test set, the accuracy of our ML is larger than 90\% as discussed below, indicating a very good model.
In order to demonstrate the transferability of the machine learning descriptor, the test set used to verify the accuracy of the predictor is constructed from a different database that has recently obtained 54 2D magnetic ordered materials from bulk compounds that are potentially exfoliable~\cite{Mounet2018}.
Finally, in the training process we consider only compounds with extreme behavior, that is, those that are clearly defined within the limits of the AFM and FM ordering. 

\textit{Numerical representation of magnetic 2D compounds:}
The second component of machine learning models involves the mapping of the material-to-attribute connection, i.e., a computer-friendly representation of the materials. These numerical representations of materials include a generalized representation that comprises atomic properties along with structural features \cite{Ward2016} and also complex representations compressing electronic properties \cite{Isayev2015}, e.g., electronic density.
In order to find trends in the chemical space that do not depend on complex features that require \textit{a priori} knowledge of the electronic properties, we use features associated with the properties of the constituent atoms of the compound. In Table \ref{tab:features}, the primary atomic features are declared.
We discriminated atomic features related to the cations and anions by the sub-indexes $c$ and $a$, respectively.
The 2D compounds in the original database are usually formed by a unique transition metal element (i.e., cation) and different anions (or only one element as an anion). When there is more than one element as anion we consider statistical functions of the atomic features \cite{Ward2016} as the primary features in the  ML classification. For a material formed by $n_{s}$ anions and $N$ atoms in the unit cell, we introduce for each atomic property $\gamma$ the statistical function $\bar{\gamma}=\left(\sum^{n_{s}}_{i=1}\gamma_{i}/\right)n_{s}$, which stand for the average value of the property.
This redefinition of the input features allows unifying the size of the feature space for compounds with a different number of atoms, which can represent a barrier to simultaneously classify binary and ternary 2D compounds.

\begin{table}[h!]
    \centering
    \begin{tabular}{cp{6cm}}
    Property & Description \\ \hline
    $Z$     & Atomic number \\ 
    $\chi$ & Pauling electronegativity \\
    $\mathcal{G}$ & Periodic group \\ 
    $v$ & Valence \\ 
    \textit{\o} & Unfilled valence orbitals \\ 
    $\mathcal{E}$, $\mathcal{I}$ & Electron affinity and ionization potential\\ 
    $\epsilon^{ho}$, $\epsilon^{lu}$ & Highest-occupied and lowest-unoccupied Kohn-Sham eigenvalue \\
    $\alpha$ & Polarizability \\ 
    $r$ & Atomic non-bonded radius \\
    $r_v$ & Radius of the last occupied valence orbital \\
    $r_{s}$, $r_{p}$ & Extensions  of the $s$ and $p$ orbitals$^*$ \\
    \hline
    \multicolumn{2}{p{0.45\textwidth}}{$^*$ i.e., the radii where the radial probability density of the valence $s$ and $p$ orbitals are maximal.}
    \end{tabular}
    \caption{Atomic and structural properties included for the construction of primary features by the $\gamma$ statistical operations.}
    \label{tab:features}
\end{table}


\begin{figure*}[h!]
    \centering
\includegraphics[width=15cm]{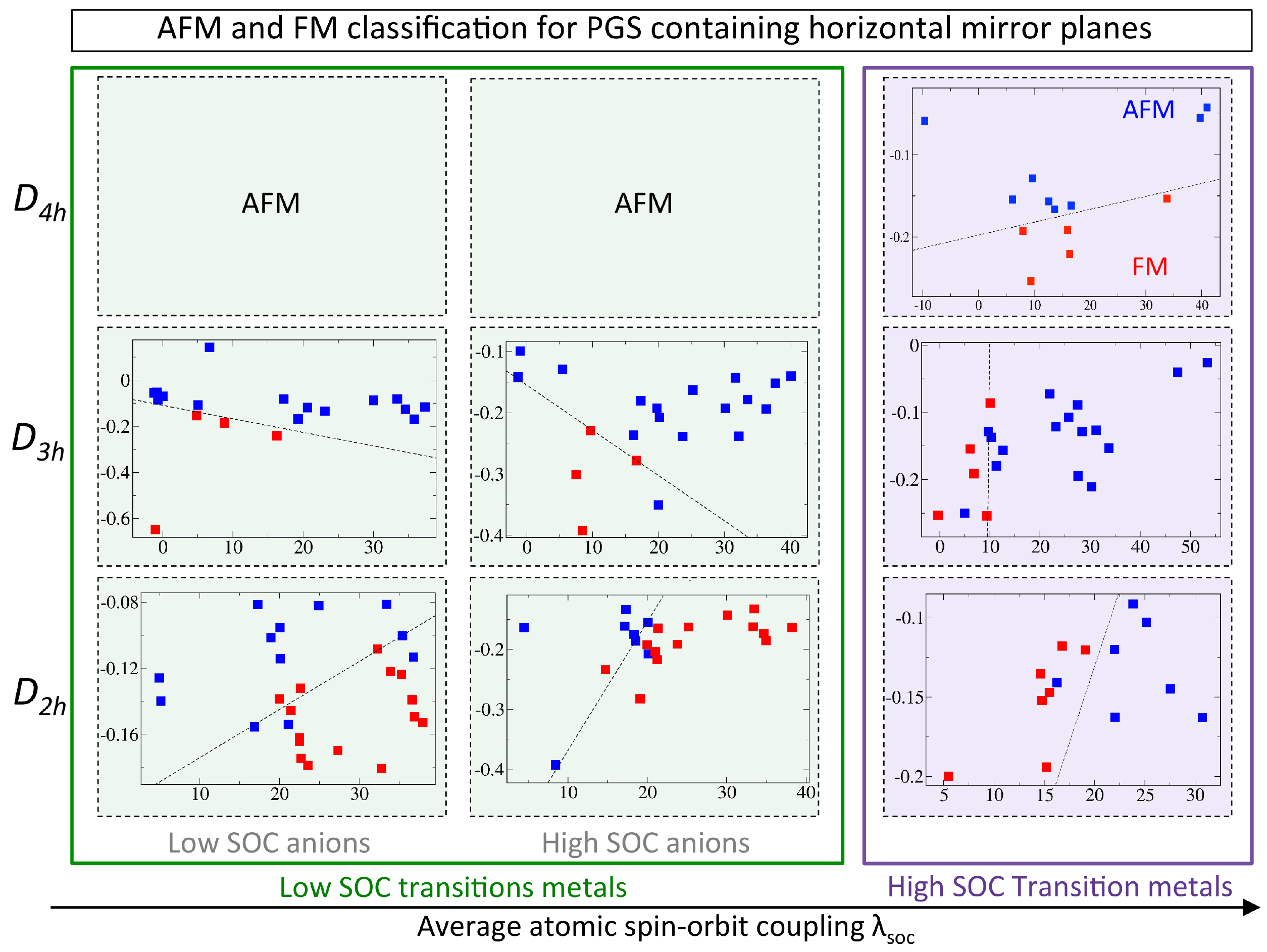} 
    \caption{Machine learning classification separating AFM (blue) from FM (red) for compounds with crystal point groups containing horizontal planes, $D_{nh}$ with $n=$ 2, 3, and 4. The descriptors $\mathcal{D}^{\sigma_h}_{x}$ ($x$ axis) and $\mathcal{D}^{\sigma_h}_{x}$ ($y$ axis) simultaneously discriminate magnetic ordering  in compound with different crystal space groups. Green and purple regions stand for the classification for compounds formed by TM with low-SOC and high-SOC, respectively. The $x$ axis stand for average atomic SOC of TM and anions.}
    \label{fig:Class1_AFMvsFM}
\end{figure*}

\textit{Learning algorithm}:
For the third component of machine learning, i.e., the learning algorithms and their optimization, we adopt the sure independence screening and sparsifying operator (SISSO) approach \cite{Ouyang2018}.
This method first automatically constructs a feature space and subsequently selects the feature that better separates the classes in the training set, e.g., AFM and FM. Based on the initial set of primary features $\Phi_0$, including the statistical combination of atomic properties (Table~\ref{tab:features}) for anions, analytical operations are used to combinatorially associate primary features with the same unit. 
The use of the primary features listed in Table~\ref{tab:features} (i.e., atomic properties) has proved to be a powerful way to do machine learning in materials. This kind of analysis has been used to predict the stability of 2D materials~\cite{Schleder2020_ACS}, topological phases~\cite{Cao2020}, and structural configurations~\cite{Ghiringhelli2015,Ghiringhelli2017}. Additionally, this set of features, in general, represented the main trends in the chemical space for atomic properties~\cite{jolly1991modern,cotton1999advanced,petrucci2007general}. 
The regions for FM and AFM classes are defined as convex hull. The descriptor, i.e., the combination of atomic features that minimizes the overlap region between convex hulls, is selected by systematically computing all overlap regions defined in a two-dimensional map, where each axis is given by a specific function of atomic properties.  

In the training set, there are compounds with the same composition and even stoichiometry that have \textit{different magnetic ordering}. These compounds are undistinguished in the feature space 
defined only by atomic properties. However, even including structural features for the representation of the materials, the random forest algorithm used in the previous section provides great inaccuracy in classifying the materials with AFM configuration. This unsuccessful capture of trends in the magnetic ordering is similar to the reported in previous works. To overcome the intrinsic dependence of the magnetism with respect to the symmetry, we use a multi-task version of the SISSO method~\cite{Ouyang2019}. Specifically, in this novel approach, a machine learning descriptor classifying the FM and AFM classes is simultaneously optimized for different groups of compounds (i.e., tasks) that contain both classes. Here, these groups are all symmetry prototypes in the C2DB. This process arrives at the best nonlinear features employed in linear models that can differentiate between classes simultaneously for every symmetry.


\textit{Trends for AFM vs FM 2D compounds:}
For AFM vs FM classifications based on the multi-task SISSO method, we evaluate different ways to group materials that have simultaneously the same descriptor. Each class (or group) of 2D materials is divided into AFM and FM ordering. Since these magnetic configurations can strongly depend on the symmetry, one can expect that the patterns in the chemical space for AFM and FM are different for compounds with different point groups symmetry classes (e.g., polar, non-polar, chiral, non-chiral, centrosymmetric). Thus, materials are grouped according to the group symmetry classes and the symmetry operation contained in the crystal point group symmetry (CPGS). In principle, the multi-task process can potentially capture the relation between the crystal symmetry and the magnetic ordering, or at least, identify patterns that are different across the PGS. 
For instance, we consider three different strategies to construct classes of 2D compounds containing AFM and FM ordering, namely, according to: i) the polarity and chirality of the CPGS, ii) the larger order of rotation symmetries in the crystal point groups (e.g., compounds with symmetry axis $R_3$, $R_4$, and $R_6$ form different classes of 2D compounds), and iii) the type of mirror symmetry planes in the CPGS (e.g., compounds with CPGS containing horizontal ${(\sigma_h)}$, diagonal ${(\sigma_d)}$ or vertical  ${(\sigma_v)}$ mirror planes form different groups). 
We note that the 
grouping of compounds based on symmetry operations (groups $i-ii$) leads to a limited SISSO model accuracy of about 80\%, which is larger than the accuracy of 70\% of previous random forest models.

In order to increase the classification efficiency, separations based on materials properties are also considered. Specifically, 2D compounds are also grouped according to materials properties such as the band gap, bulk spin-orbit coupling (i.e., calculated as the average of the atomic spin-orbit coupling), and the existence of either 3$d$, 4$d$, or 5$d$ transition metals. These materials' properties are selected based on the magnetic physical mechanism in three-dimensional compounds, as discussed in the next section. 

We find that the SOC is a determinant factor for high accuracy in the classifier separating AFM from FM in a materials map.
In Figure~\ref{fig:Class1_AFMvsFM}, the classification model for groups formed by the interpolation of crystalline SOC and different types of mirror symmetry planes is presented. For each group of materials, AFM and FM compounds are illustrated in red and blue, respectively. For instance, in Figure~\ref{fig:Class1_AFMvsFM} compounds with point groups symmetry containing at least one horizontal mirror plane (i.e., $D_{2h}$, $D_{3h}$, $D_{4h}$, and $D_{6h}$) are used to train the multi-task ML classifier. The SOC effect is introduced by dividing compounds into those containing TM with high SOC (i.e., 4$d$ and 5$d$ atoms) and those with low-SOC (i.e., 3$d$ atoms), which are illustrated in purple and green regions, respectively. Compounds formed by low-SOC TM are also divided in terms of the atomic SOC of anions as 2D compounds with low-SOC and high-SOC anions (Figure~\ref{fig:Class1_AFMvsFM}). 
The axes determining the predicted material map are given by the two-dimensional descriptor $\vec{\mathcal{D}}^{(\sigma_h)}$ with the components $\mathcal{D}^{\sigma_h}_{x}$ and $\mathcal{D}^{\sigma_h}_{y}$ given by the functional mathematical form: 
\begin{equation}
\mathcal{D}^{\sigma_h}_{x}=\left({\epsilon^{lu}_c+\epsilon^{lu}_a}\right)\mathcal{G}_{c}/r_{p,c}
\end{equation}
$$\mathcal{D}^{\sigma_h}_{y}=r_{v,a}/r_{c}\left( \mathcal{E}_{c}+\mathcal{E}_{a}\right),$$
where the superindex ${\sigma_h}$ stand for horizontal mirror planes,  $\epsilon^{lu}_c$ and $\epsilon^{lu}_a$ are the lowest-occupied Kohn-Sham eigenvalues for the cations (i.e., TMs) and anions, $\mathcal{G}_c$ is the periodic group of the cation, $r_{p,c}$ is the radii for the $p$-orbitals of the cations, $r_{v,a}$ is the radius of the last occupied valence orbital for anions, $r_c$ is the atomic non-bonded radius for cations, and $\mathcal{E}_{c,a}$ are the electron affinity for anions and cations. In the training set, the accuracy of the SISSO descriptor is larger than 98\%, while the accuracy in the test set is larger than 90\%. 
Importantly, even with the same descriptor $\vec{\mathcal{D}}$, the tendency for every symmetry is almost unique, i.e., the boundary line defining AFM and FM regions in the materials maps (dashed lines in Figure~\ref{fig:Class1_AFMvsFM}) is different for each map. 
For instance, for the crystal point group $D_{2h}$, the slope of the boundary line is positive regardless of the atomic SOC of the constituent atoms (Figure~\ref{fig:Class1_AFMvsFM}), but its position in the materials map changes as the atomic SOC changes. In contrast, for the CPGS $D_{3h}$, the slope of the boundary lines drastically increases when the average atomic SOC increases. Also, for compounds with CPGS $D_{4h}$, a low average atomic SOC directly indicates an AFM phase, with a clear demarcated tendency to forbid the FM order.


Besides the ML classifier for compounds with CPGS containing horizontal mirror planes, we also find a functional rule $\vec{\mathcal{D}}^{\sigma_v}$ leading to a separation between AFM and FM in compounds with CPGS containing mirror vertical planes. The component of the two-dimensional descriptor $\vec{\mathcal{D}}^{\sigma_v}$ are given by: 
\begin{equation}
\mathcal{D}^{\sigma_v}_{x}=\chi^{2}_{c}\mathcal{G}_{c}/\mathcal{E}_{a}
\end{equation}
$$\mathcal{D}^{\sigma_v}_{y}=\mathcal{G}_{c}v_{c}\left(\mathcal{G}_{a}+v_{c}\right),$$
where $\chi_c$ is the Pauli electronegativity of TMs, $\mathcal{G}_{c,a}$ are the periodic group for cations and anions, $\mathcal{E}_{a}$ is the electron affinity of anions, and $v_c$ is the valence of the cation. The descriptor $\vec{\mathcal{D}}^{\sigma_v}$ provides a reasonable good classification with an accuracy of 92\% in the training set and 87\% in the test set. This accuracy is significantly higher than that of  70\% in previous machine learning models~\cite{Kabiraj2020} for magnetic ordering. Although we are mainly motivated by obtaining a rule that allows us to classify 2D materials into AFM and FM, rather than interpreting a physical mechanism, one can find some relationships between atomic properties that could reveal in the future the physical mechanism determining the magnetic ground state. 

\subsection*{Theoretical trends and captured mechanism}

\begin{figure}[h!]
    \centering
    \includegraphics[width=\linewidth]{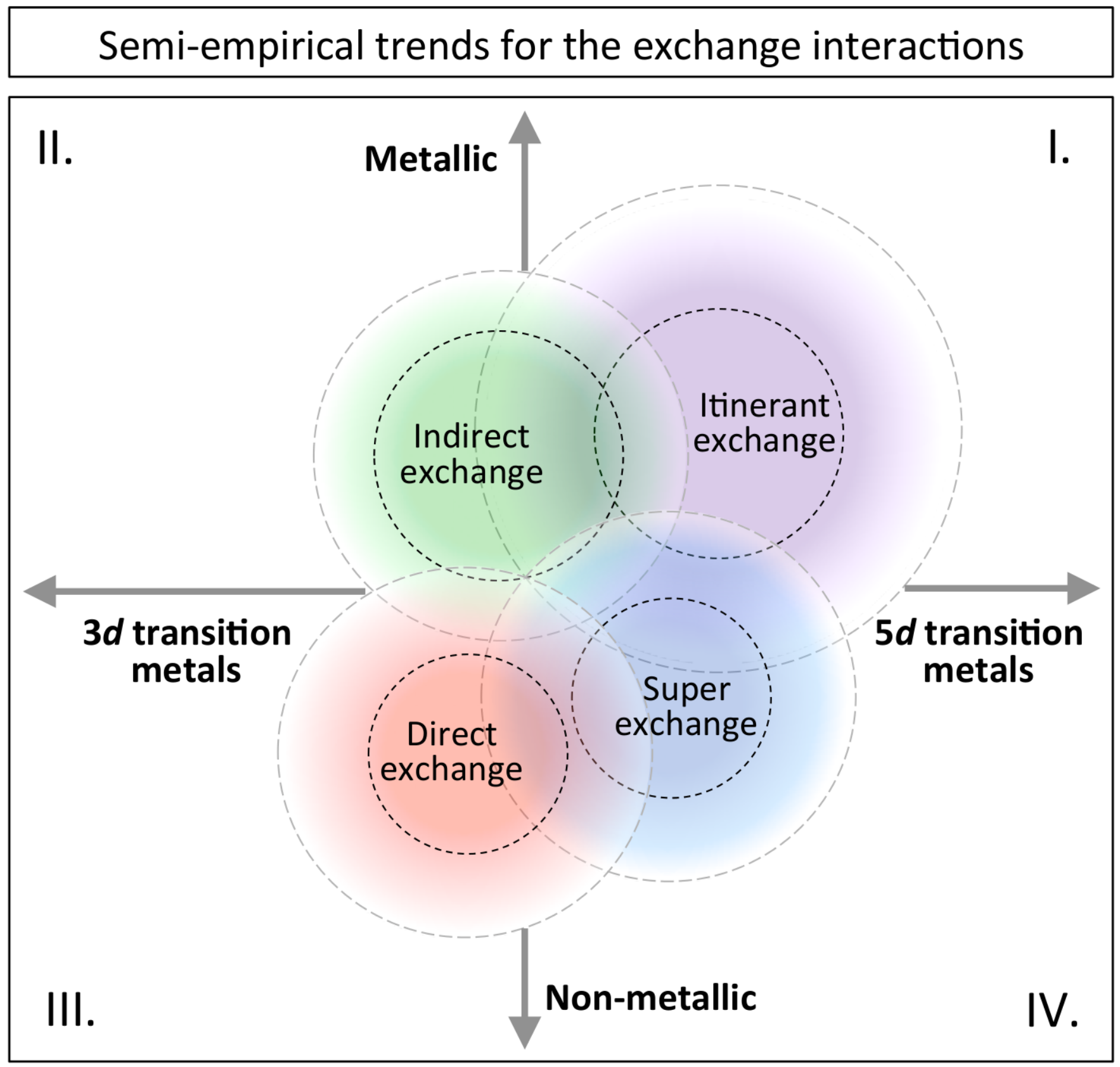}
    \caption{Materials map for exchange interaction in bulk materials\cite{magnet_book}. The four quadrants I-IV correspond to the indirect exchange region (green), itinerant exchange region (purple), super exchange region (blue), and direct exchange region (red), respectively. }
    \label{fig:class_map}
\end{figure}

In bulk magnetic materials, the effective spin magnetic moment is usually different from zero due to the presence of ions with partially filled electronic $d$- (transition metals) or $f$-shells (rare earths). In this case, Hund’s rules lead to a finite spin (S) and orbital (L) magnetic moments - the total magnetic moment is J = S + L. Usually, the orbital magnetic moments (L) are quenched for transition metal elements and only spin magnetic moments (S) are considered. When the magnetic ions have completely filled (or empty) electronic $d$ or $f$-shell, the effective spin magnetic moment is zero leading to diamagnetic property — the so-called non-magnetic materials. Molecular field theory can explain the observation of a net magnetization due to the competition between the exchange interactions among magnetic ions leading to a macroscopic ordering of the system and the crystal field splitting of $d$-orbitals of TMs. The analogy with three-dimensional compounds paves the route for the interpretation of the obtained ML models for the existence of magnetism and classification of the magnetic ordering.  To do so, the materials maps for magnetic bulk materials is briefly discussed (see Figure \ref{fig:class_map}). 

Notwithstanding its inherent complexity, there are clear trends for the magnetic order in bulk materials, as illustrated in Figure \ref{fig:class_map}. These trends  
are represented as a material map in the feature space defined by the metallic behavior of the compounds and the localization of the atomic magnetic moments~\cite{magnet_book}. Four quadrants can thus be identified in Figure \ref{fig:class_map}: (I.)  metallic compounds formed by 5$d$ TM, where the itinerant exchange tends to be predominant; (II.) In metallic compounds formed by 3$d$ TM, the magnetism is mainly mediated by the indirect exchange; (III.) For non-metallic materials formed by 3$d$ TM, the direct exchange is typically dominant; and (IV.) For non-metallic with 5$d$ TM, the super exchange usually explains the magnetic ground state. This materials map also reveals a trend in magnetic ordering that separates FM materials in quadrants I and II, while locating AFM materials in quadrants III and IV. Naturally, this is not a general rule, since in practice the symmetry of the crystal plays a role in the description of the magnetic state and there is no atomic composition that \textit{a priori} dictates a specific exchange interaction. The description of magnetic order in 2D materials is even more complex.

One of the bottlenecks for understanding the trends of magnetism in 2D materials, and then proposing functional applications with them, is due to only a few experimental realizations and their contradiction with the materials map for Bulk compounds (Figure \ref{fig:class_map}).
This includes the more recent discovery of room-temperature ferromagnetism in monolayer VSe$_2$~\cite{Bonilla2018} and MnSe$_2$~\cite{OHara2018} stabilized by the strong magnetocrystalline anisotropy, which suppresses the Mermin–Wanger restriction. Both compounds, as well as the reported strong out-of-plane magnetic anisotropy in Fe$_3$GeTe~\cite{Fei2018}, have itinerant ferromagnetism. These compounds belong to the purple region of the materials map in Figure~\ref{fig:class_map}, but it contradicts the illustrated trend since they are formed by 3d TM. In contrast, the strong direct ferromagnetic exchange (red region in Figure~\ref{fig:class_map}) is modeled as an Ising-type magnetism, which describes the AFM with an ordering temperature of 118 K in FePS$_3$~\cite{Lee2016}.  
Additionally, the interlayer magnetic coupling in bilayers of 2D materials can give rise to novel physical properties~\cite{Sivadas2018,Cardoso2018,Su_rez_Morell_2019,Jiang2018,Klein1218,Kim_2019,Kim11131,Kim2019}, which can not be directly accommodated in the materials map of Figure \ref{fig:class_map}. However, the ML analysis of theoretical predicted and hypothetical compounds in previous sections provide clear trends that are actually related to a theoretical explanation for magnetism in 2D compounds, as discussed below. 

\textit{Captured mechanism for the existence of magnetism:}
\begin{figure}[h!]
    \centering
    \includegraphics[width=7cm]{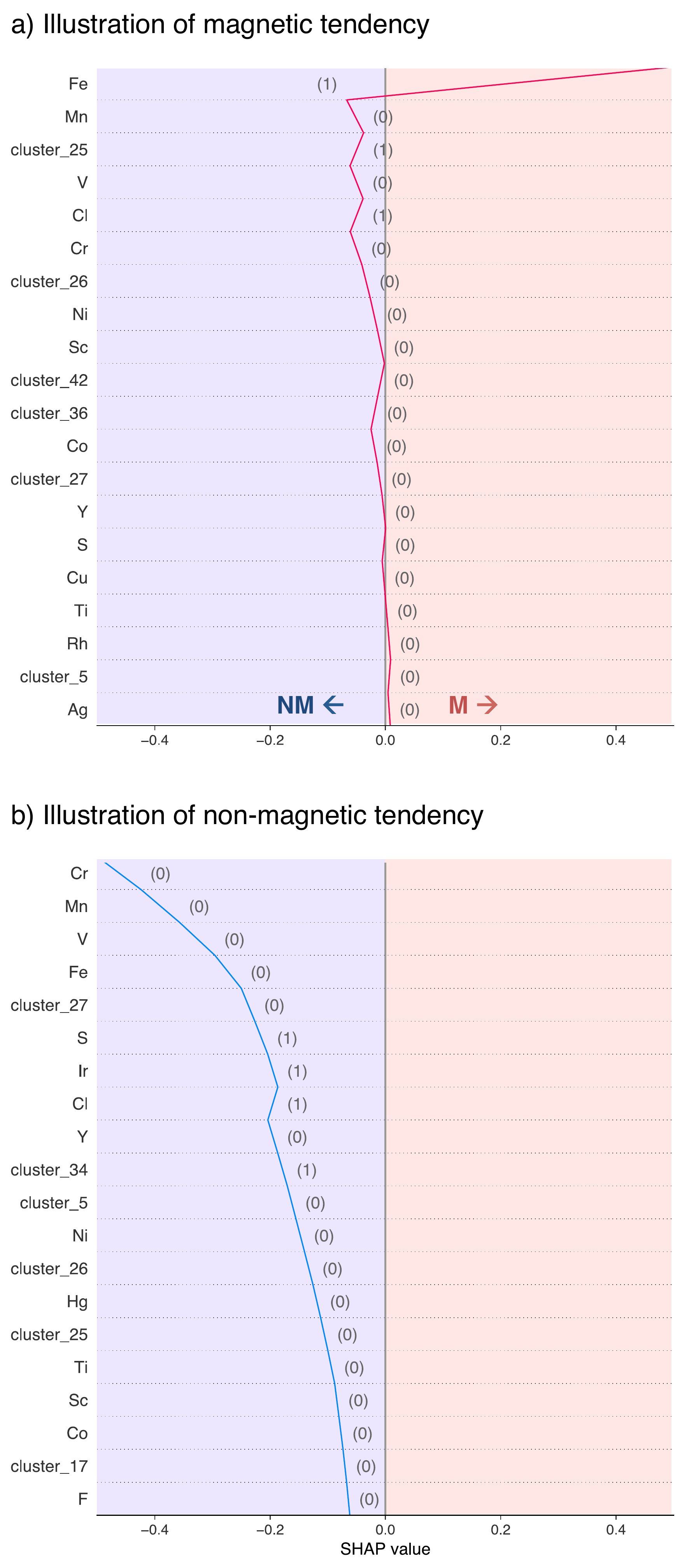}
    \caption{SHAP decision plots of two distinct materials showing how specific compositional/structural features values (0: False; 1: True) contribute to the model classification (NM: left; M: right). Here it is shown only the material-specific top-20 features with respect to the absolute SHAP values; a) \ce{FeCl2} in the {\it cluster 25} crystal structure (TMDC-H): The presence of Fe alone shows to be sufficient for the M classification in this case; b) \ce{IrClS} in the {\it cluster 34} crystal structure (TMHC): The absence of many 3d TM in the composition, denoted by 0, contributes to an NM classification.}
    \label{fig:Shap_exemplo}
\end{figure}
To better understand the physical mechanism captured by our ML model, we first illustrate how the prediction is constructed based on the SHAP scheme. For two different 2D materials, figure~\ref{fig:Shap_exemplo} shows the resulting total weight defining a non-magnetic and magnetic tendency, which are represented by negative values (left side in blue) and positive values (right side in red), respectively. Note that both the existence (represented by 1) and the absence (represented by 0) of a given structure or chemical species contribute to defining the existence of magnetism. For instance, in the first two lines of Figure~\ref{fig:Shap_exemplo}a, one can see that the existence of Fe atoms in the composition of the compound positively contributes to the classification of the compound as magnetic, but the absence of Mn atoms negatively contributes to the classification as magnetic (i.e., the contribution to the total weight is negative, and the absence of Mn tends to make the compounds non-magnetic). Another important consideration in these trends is that, if the existence of a specific atomic specie contributes with a positive weight $\xi_X$ to the magnetism, it does not imply that the contribution of the absence of X is $-\xi_X$. For instance, in Figure~\ref{fig:Shap_exemplo}a, the existence of Fe atoms leads to high positive weight, however, in Figure~\ref{fig:Shap_exemplo}b (fourth line), the absence of the same atom has a small negative weight. This means that the total weight determining if a given compound is magnetic or non-magnetic is not defined by a few atoms in the composition or a specific structure, but by the collective weights of the absence and existence of atoms and structural prototypes (i.e., symmetries and Wyckoff positions). 

The random forest and SHAP models indicated that the existence of 3$d$ TMs strongly contributes with a positive weight to the existence of magnetism (Figure~\ref{fig:TM_shap}), where the V, Cr, Mn, and Fe have the largest positive contributions. In contrast, besides the existence of Y atoms that positively contribute magnetism, 4$d$ and 5$d$ TMs do not have a predominant effect in the existence or absence of magnetism, having then a minor, but non-zero contribution. For instance, while Rh and Hg atoms have a small positive contribution to the existence of magnetism, Ir atoms have a relatively small contribution to the absence of magnetism. 

\begin{figure}[h!]
    \centering
    \includegraphics[width=\linewidth]{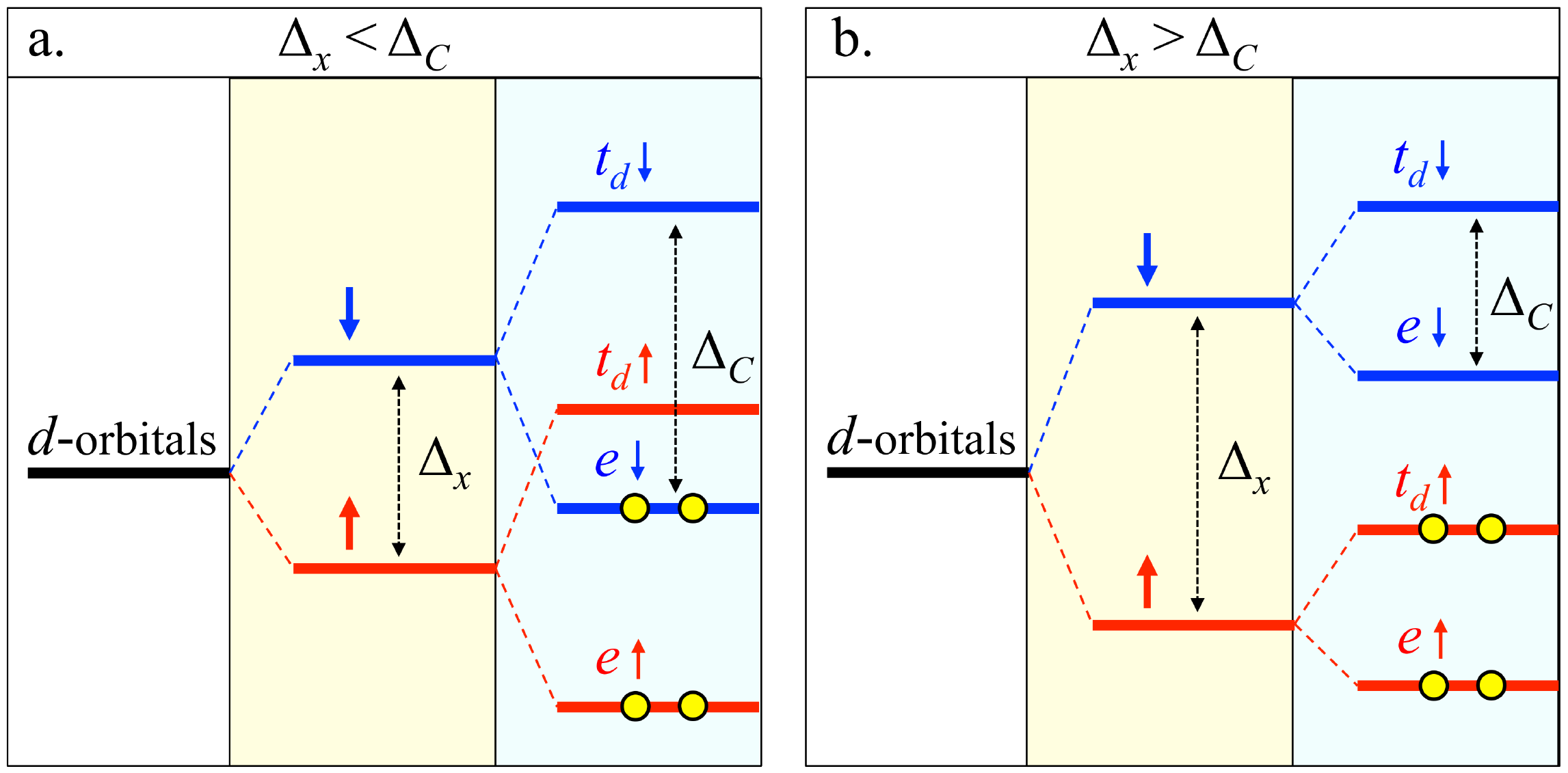}
    \caption{Schematic representation of the d-orbitals splitting including the exchange splitting $\Delta_x$ (yellow region) and the crystal field splitting $\Delta_c$ (blue region). a) the case where $\Delta_x$ < $\Delta_c$. b) the case where $\Delta_x$ > $\Delta_c$. Situations a) and b) illustrate respectively a low-spin configuration and a high spin configuration (yellow dots indicate electrons)
    }
    \label{fig:Spin_conf}
\end{figure}

These trends can be interpreted in terms of the competition between the exchange splitting (ES) $\Delta_x$ and the crystal field splitting (CFS) $\Delta_c$. 
In figure~\ref{fig:Spin_conf}, the successive effect of $\Delta_x$ and $\Delta_c$ are represented for the case of $d$-orbitals. The introduction of the exchange splitting separates spin up from spin-down; subsequently, the symmetry-induced crystal field splitting further separates these levels. As an example, the T$_d$ crystal field effect is shown in Figure~\ref{fig:Spin_conf}, which splits the five-degenerated $d$ levels into $t_2$ and $e$ levels. We can draw two different scenarios, namely: $\Delta_x$ < $\Delta_c$ (Figure~\ref{fig:Spin_conf}a) and $\Delta_x$ > $\Delta_c$ (Figure~\ref{fig:Spin_conf}b). These two situations can be directly associated with the type of $d$-orbitals. Specifically, 3$d$-orbitals are usually related to relatively small crystal fields, while 4$d$-orbitals and 5$d$-orbitals typically lead to large crystal field splitting. Thus, the connection between the $d$-orbital type and the existence of magnetization has an intrinsic correlation to the competition among CFS and ES.
Specifically, the local magnetic moment depends on the number of electrons; however, a simple empirical model based on $\Delta_x$ and $\Delta_c$ usually gives a reasonable intuition of the magnetization. For instance, for $\Delta_x$ < $\Delta_c$ (Figure~\ref{fig:Spin_conf}a), both spin up (red) and down (blue) can be simultaneously populated, favoring a spin configuration with zero or relatively small local magnetic moment (i.e., compound tending to have zero magnetization) or having a relatively small magnetic moment. In contrast, when  $\Delta_x$ > $\Delta_c$ (Figure~\ref{fig:Spin_conf}b), only states of one spin type are populated. 
Consequently, the spin configuration tends to have non-zero (and relatively large) local magnetic moment (i.e., FM or AFM compounds). In this last scenario, the final magnetic configuration depends on other factors (e.g., the SOC), as we analyze in the next section. Although the schematic model illustrated in Figure~\ref{fig:Spin_conf} focused on the $T_d$ symmetry, the conclusions are valid for any other general crystal field splitting since the physical mechanism depends on the order of the levels and not on the specific final symmetry.
These empirical observations are in agreement with the trends captured by our model: \textit{i}) 4$d$- and 5$d$ orbitals tend to be delocalized, which results in a relatively large CFS ($\Delta_x$ < $\Delta_c$) and hence, a local magnetic configuration with small or zero local magnetic moments, and \textit{ii}) 3$d$-orbital have a strong localization (compared to 4$d$- and 5$d$-orbitals) and consequently, relatively smaller CFS ($\Delta_x$ > $\Delta_c$), leading then to a magnetic configuration with large local magnetic moments.  
Summarizing, our ML model captures the relation between 3$d$-orbitals (large values in the SHAP plot of Figure~\ref{fig:TM_shap}) and the fact that 4$d$- and 5$d$-orbitals) cannot directly be related to the absence or existence of magnetism (near zero values in Figure~\ref{fig:TM_shap}). 

The distinction of magnetic and non-magnetic 2D compounds is based on the existence and absence of specific atoms, as well as specific symmetries, which, naturally, is intrinsically related to the atomic properties (e.g., unpaired electrons, exchange strength, bond length, coordination number) and symmetry point groups. We have two types of features: structural and atomic. Structural features can directly correlate to symmetry, coordination number and exchange strength. On the other hand, atomic features correlate to the number of unpaired electrons, exchange strength and bond length. At the end, we are able to probe all important features for magnetism. Specifically, we find that 2D compounds with non-zero local magnetic moments are typically formed by 3d transition metals, which are mainly differentiated with respect to 4d and 5d by the high orbital-localization degree and relatively small atomic radii. These atomic features result in a high exchange splitting~\cite{Nandy2020}, which as described in Fig.~\ref{fig:Spin_conf}, leads to a high spin configuration or in other words, a tendency to be magnetic.

The previously noted relation between atomic orbital localization and local magnetic moments in 2D materials can also be used to extract information from the SHAP plots for anions (Figure~\ref{fig:anions_shap}). Our model shows that the existence of halides in a compound implies a positive weight for the existence of magnetization. Indeed, we verify that the $p$-orbitals in halides tend to be more localized than chalcogen $p$-orbitals. It is then expected that Halides also lead to relatively small CFS ($\Delta_x$ > $\Delta_c$), and hence, a non-zero local magnetic moment.  

We also find that the structure plays an important role in the existence of magnetism in two-dimensional materials. By analyzing the structural cluster which contributes to an M classification, it is possible to delineate that the regularity of transition metals in the crystal structure directly reflects on the exchange interactions and, thus, favors the emergence of magnetic order. While crystal structures with regular TM sublattices highly favor magnetism, structures with chemical or structural irregularities seem to disturb the long-range interactions between ions. However, the ground state magnetic configuration depends on other features, as captured by the ML model and explained in the next section.

\textit{Captured mechanism for magnetic ordering:}
As previously discussed, we find that in 2D-materials, the SOC is a determinant factor for the classification accuracy of AFM and FM compounds. This is a strong result relating the patterns of the magnetic ordering classification with the SOC strength, i.e., when the atomic average of the SOC for the constituent elements increases in the 2D compounds, the distribution of AFM and FM compounds change in material maps. Previous calculations indicate that as the atomic combination forming the compounds has larger SOC, the magnetic anisotropy increases, which is fundamental to allow 2D magnetism~\cite{2017Lado,Jeongwoo2019}. The ML SISSO classification (see Figure~\ref{fig:Class1_AFMvsFM} for the specific case of CPGS $D_nh$ with $n=2, 3, 4$ and 6) indicates that material maps and boundaries defining AFM and FM regions can strongly depend on the magnetic anisotropy. On the other hand, our ML model also indicates that the magnetic order configuration depends on the symmetry, i.e., higher symmetries favor the AFM order. For instance, compounds with more symmetric PGS tend to be AFM (see PGS $D_{4h}$ in Figure~\ref{fig:Class1_AFMvsFM}). As the PGS have lower symmetry, the proportion of FM compounds with respect to AFM increases.   

Even though  SISSO  is a very powerful ML model, it is certainly not the best-suited model to interpret the results. The descriptors that are built can be complex and do not have a direct interpretation. However, some hints can appear in the features defining the descriptors. Indeed, equations 1 and 2 give us some hints of what is important to classify a compound as FM or AFM. Compounds formed by atoms with large (small) valence tend to have AFM (FM) ordering, and compounds with a large (small) $r_{pM}$ tend to be FM (AFM). In the opposite way, for the space group P-3m1, compounds formed by atoms with large (small) lowest-occupied Kohn-Sham level tend to have AFM (FM) ordering. Although the interpretation of this descriptor is not obvious, one can infer that the value of the radii of the $p$-orbitals for cations is related to the size of the atoms that separate the transition metals. When the size of TM increases (with the symmetry as a constant), the compound tends to be FM.
Similarly, the descriptor also reveals that the classification of AFM and FM ordering depends on the electron affinity, which is related to how the electrons are transferred in atoms to form atomic bonds in the 2D compound.  

Additionally, for $\vec{\mathcal{D}}^{\sigma_v}$, the valence orbitals $v$, the Pauling electronegativity, and the periodic groups appear to play the main role in defining the magnetic ordering. Indeed, the simplest interpretation of the exchange interaction is based on the unfilled valence orbitals and Pauli electronegativity. There is no direct relation between these atomic features and the physical mechanism defining the magnetic ordering. However, with the predicted material maps, one can infer the magnetic ordering by means of the atomic composition in a defined symmetry. When the considered symmetry changes, the classification can also be different, which illustrates the robustness of the method and determined descriptors.



\subsection*{Machine learning model validation and potential applications}

In this section, we use the proposed scheme to predict the existence of magnetism and magnetic structure of novel 2D materials candidates, i.e., compounds that are not included in the initial database used to train and test the ML model. The ML prediction is followed by a theoretical total energy calculation of the magnetic ground states based on \textit{ab inition} theory. DFT+SOC calculations are performed using the Vienna Ab initio Simulation Package (VASP) with the projector-augmented wave (PAW) method  \cite{vasp1, vasp2} and GGA-PBE \cite{PBE} parametrization for the exchange-correlation functional. As an illustrative example, we describe here the ML prediction and DFT verification for compounds with $D_{3h}$ PGS. The complete list of novel predicted compounds for all symmetries can be found in the Supporting Information. The ML prediction process starts with the generation of novel compounds based on trends for 2D magnetism defined in the features space formed by chemical composition and structural clusters. Afterward, we make the proposed materials go through steps of the prediction scheme I (classification for magnetic vs non-magnetic materials) and II (classification for AFM vs FM ordering in 2D compounds).

\textit{Generation of novel magnetic candidates}: In previous sections, we learned that compounds containing 3$d$-orbitals TM and halides tend to host 2D magnetism. the construction of atomic combinations is based on the selection of 3$d$ TMs V, Mn, Cr, Fe, Ni, Co, and Ti (the highest positive weight) and as anions, the halides Cl, Br, and I. These atomic species lead to 21 atomic combinations, which can be accommodated in the 51 structural prototypes (i.e., 1071 compounds). Since the weight of the atomic composition is usually larger in determining the existence of magnetism, all structural prototypes are considered for generating 2D magnetic candidates. These compounds intrinsically account for different stoichiometries and compositions (e.g., AB, AB$_2$, and A$_2$B$_3$). The hypothetical generated materials are always in the green region of the materials maps classifying AFM and FM materials since they are only formed by low SOC TMs. 

\begin{figure}[h!]
    \centering
    \includegraphics[width=\linewidth]{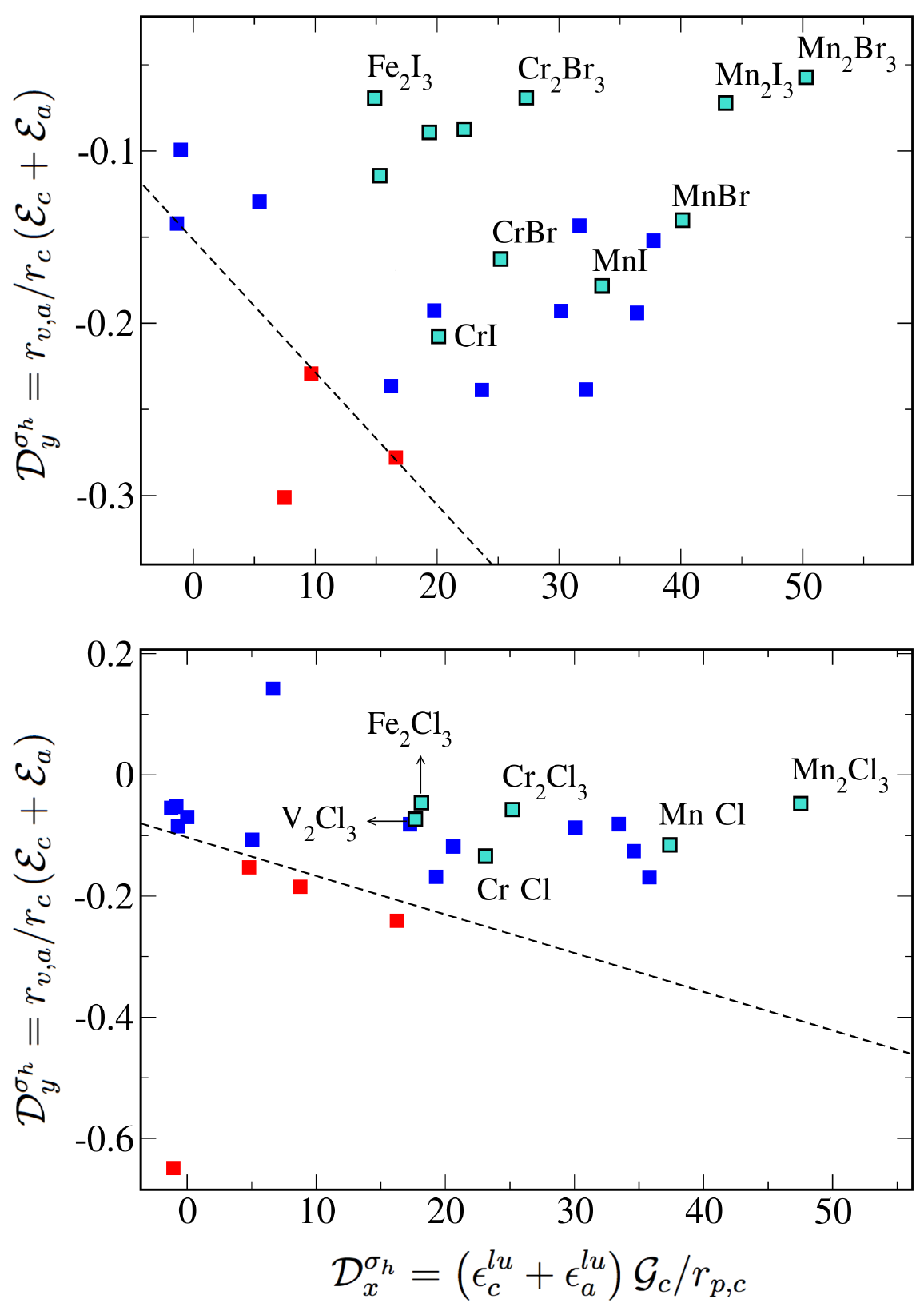}
    \caption{Prediction of novel candidates with $D_{3h}$ crystal point group symmetry with low-SOC (top) and medium-SOC (bottom). The red and blue squares stand for the ferromagnetic and antiferromagnetic compounds, respectively. Novel 2D magnetic candidates in the $d_{3h}$ point group are represented in cyan. }
    \label{fig:pred}
\end{figure}

\textit{ML model applied to novel candidates}: The list of generated compounds is compared with the initial list of compounds contained in the C2DB database and only a novel combination of ACS attributes is considered. Using the random forest model, compounds are classified as magnetic and non-magnetic. After considering compounds with a 100\% of probability to be magnetic (i.e. classified as M by 100\% of the RF's decision trees), the filtering process leads 169 novel 2D magnetic candidates. For these 169 prototypes, we use the SISSO models to identify the magnetic ordering, which results in 137 AFM and 32 FM 2D compounds. This prediction is performed by identifying the position of the compounds in the materials maps by means of the calculation of the functional form of the descriptor in terms of the atomic properties. For instance, Figure~\ref{fig:pred} show the AFM (blue), FM (red), and novel 2D magnets (cyan) the CPGS $D_{3h}$. The boundaries (dashed lines in Figure~\ref{fig:pred}) are  defined by the SISSO model in Figure~\ref{fig:Class1_AFMvsFM}. Here, 
The position of the novel candidates indicates that they are classified as AFM materials (Figure~\ref{fig:pred}). 

\textit{DFT verification}: Our DFT calculation for the selected compounds in the $D_{3h}$ PGS reveals that all compounds in composition A$_2$B$_3$ (crystal structure of MXenes, {\it cluster 23}) are not mechanically stable in this structure, i.e., the compound tends to dissociate. On the other hand, materials with the composition AB (crystal structure of ABBA InSe, {\it cluster 17}) are energetically stable with respect to the free atoms and solids formed by the constituent atoms. The magnetic ground states for the six remaining atomic combinations are studied by means of DFT calculations. Remarkably, we find that the compounds AB predicted as AFM  by the ML scheme in Figure~\ref{fig1} are confirmed to be AFM by PBE DFT+SOC total energy calculations, as shown in Table~\ref{tab:pred}. These AB compounds can be understood as a unit cell containing two quasi layers, where each quasi layers has one TM.
The DFT calculation indicates that the most stable AFM magnetic configuration for CrCl, CrBr, CrI, and MnCl contains only two TM atoms in the unit cell, i.e., TMs at different planes have opposite local magnetic moments along the $z$ axis. On the other hand, the AFM unit cell for MnBr and MnI contain four TMs, where subsequent TMs in the same plane have opposite local magnetic moment orientation, which is inverted from one plane to another. 

\begin{table}[h!]
    \centering
    \begin{tabular}{cccc}
 Compound & GS & $\Delta_{\text{AFM-NM}}$ & $\Delta_{\text{AFM-FM}}$ \\  \hline
    CrCl & AFM & -75 & -2\\  
 CrBr & AFM & -75  & -15 \\
 CrI & AFM & -73 & -62 \\  
 MnCl & AFM & -9  & -25 \\ 
 MnBr & AFM & -8 & -36 \\  
 MnI & AFM & -7  & -49  
    \end{tabular}
    \caption{Magnetic ground state (GS) for compounds with the composition AB in Figure~\ref{fig:pred}. For each combination, the energy difference  for the non-magnetic ($\Delta_{\text{AFM-NM}}$) and ferromagnetic ($\Delta_{\text{AFM-FM}}$) phases with respect to the AFM ground state are provided in meV/atom.  }
    \label{tab:pred}
\end{table}

\section{Conclusions}

The design and the discovery of novel functional materials have been historically sustained 
by an Edisonian trial-and-error approach. This expensive and time-consuming approach has been
recently substituted by an intelligent data-based approach. The basic idea is to rely
on computational databases of compounds, where the physical properties of thousands of materials have been calculated using quantum mechanical methods. In the current manuscript, we have used this approach to search for magnetic 2D compounds. This is a recent field of research with huge potential for 
functional applications.

Our study was divided into two steps. We have used the C2DB database as the source of information regarding structural, electronic, and magnetic properties of a large variety of synthesized and 
computer-designed compounds. These compounds were classified as non-magnetic (NM - diamagnetic), ferromagnetic (FM), and antiferromagnetic (AFM). It is well-known and established, however, that there are several other magnetic configurations that were not included in this list, including ferrimagnetic, helimagnet, and compounds with frustrated interactions. More complex than that, the AFM configurations can have a large variety of different accommodations for the spins, usually labeled as type-A, C, D, G, etc. The C2DB database only accounted for the more simple configuration, where neighboring spins are aligned in an anti-parallel configuration. Even though there is much discussion about the outcomes of this approach
\cite{gjerding2021recent}, the available data should be sufficient for our machine learning purposes. 

The random forest algorithm used in this work was able to successfully differentiate  
magnetic from non-magnetic compounds with great accuracy. Non-magnetic compounds were predicted with an accuracy
of $\approx 96$\%, while for magnetic compounds this accuracy was around 85\%. The SHAP analysis allowed us to understand why a material was classified as M or NM, indicating that 3d TMs greatly impact the classification into M in comparison to 4d and 5d TMs. Anions showed lower impact in comparison, but a qualitative analysis showed that halides contribute to magnetic order, in contrast to chalcogens.

An analysis of the crystal structure of the magnetic 2D compounds revealed that the transition metal sublattice regularity is a common characteristic in all structural clusters favoring magnetic ordering. This general trend is supported by the many experimental and theoretical studies of magnetism in monolayers in recent years.


The classification of FM and AFM compounds is more subtle than the previous one and 
demands a different approach. We have used the SISSO method which was able to successfully find
a descriptor that is able to classify FM and AFM compounds. 
We have used a large set of atomic features to classify these compounds, and 
the final descriptor included information regarding the electron affinity, the number of
valence electrons, and the electronegativity of the atoms that constitute the compound.
More importantly, our results indicate that it is necessary to separate the compounds according to their space groups and spin-orbit coupling in order to identify patterns that classify AFM and FM 2D compounds. Consequently, the magnetic order of material involves a complex competition among structural and electronic properties. 

The information provided here can be useful in the search for novel 2D magnetic materials, expanding the possibilities of this expanding and exciting area to compounds that can be more stable and easy to synthesize than the candidates observed so far. 

\begin{acknowledgement}

The authors acknowledge the National Laboratory for Scientific Computing (LNCC/MCTI, Brazil) for providing HPC resources of the SDumont supercomputer. The authors also thank the Brazilian funding agencies FAPESP (Grants 17/02317-2, 18/11856-7 and 	18/11641-0) and CNPq (309384/2020-6) for financial support. 
\end{acknowledgement}
\begin{suppinfo}
In the Supporting Information we provide extra information regarding details of the random forest classification scheme, the details about the structural labels and detailed SHAP graphs for different features. The complete list of novel 2D magnets candidates classified as AFM and FM by the ML model is also provided.  
\end{suppinfo}
\bibliography{ML_2DMag}



\section*{Supplementary material}

# Machine Learning Study of the Magnetic Ordering in 2D Materials

Carlos Mera Acosta, Elton Ogoshi, José Antonio Souza, and Gustavo M. Dalpian

Federal University of ABC (UFABC), 09210-580, Santo André, São Paulo, Brazil
carlos.acosta@ufabc.edu.br; elton.ogoshi@ufabc.edu.br; gustavo.dalpian@ufabc.edu.br

# Supporting Information

## 1. Random forest classification of magnetic and non-magnetic 2D compounds

### 1.1 Model evaluation metric

To evaluate the RF classification performance we used the out-of-bag (OOB) methodology, which consists of training and validating the model using 100% of the available data. This is done by using for validation only Decision Trees (DTs) in the RF ensemble which were not trained with the specific instance. This way, each material is validated only by the DTs that have never seen the specific material.]

As the magnetic materials (M) are a minority class (25.8%), meaning that the set of materials is class unbalanced, the most appropriate accuracy metric for a classification model is the F1 score. The F1 score takes into consideration not only the overall model accuracy but also takes into special consideration the model's accuracy of the minority class. It is defined by the harmonic mean of the precision and the recall of the M class, and it is given by the following expression:

$$F1 \;=\; 2\frac{(precision.recall)}{precision + recall} \;=\; \frac{TM}{TM + \frac{1}{2}(FM+FNM)}$$

where TM is the true M (the accuracy of the model for the M class), FM is the false M (the ratio of M materials incorrectly classified as NM), and FNM is the false NM (the ratio of NM materials incorrectly classified as M).

In the context of a highly class-unbalanced dataset with 99% of instances belonging to class A and 1% to class B, for example, a naive model would always output class A, and it would have an overall accuracy of 99%; however, its F1 score would be 0%.

## 1.2. Defining a structural label using an unsupervised approach

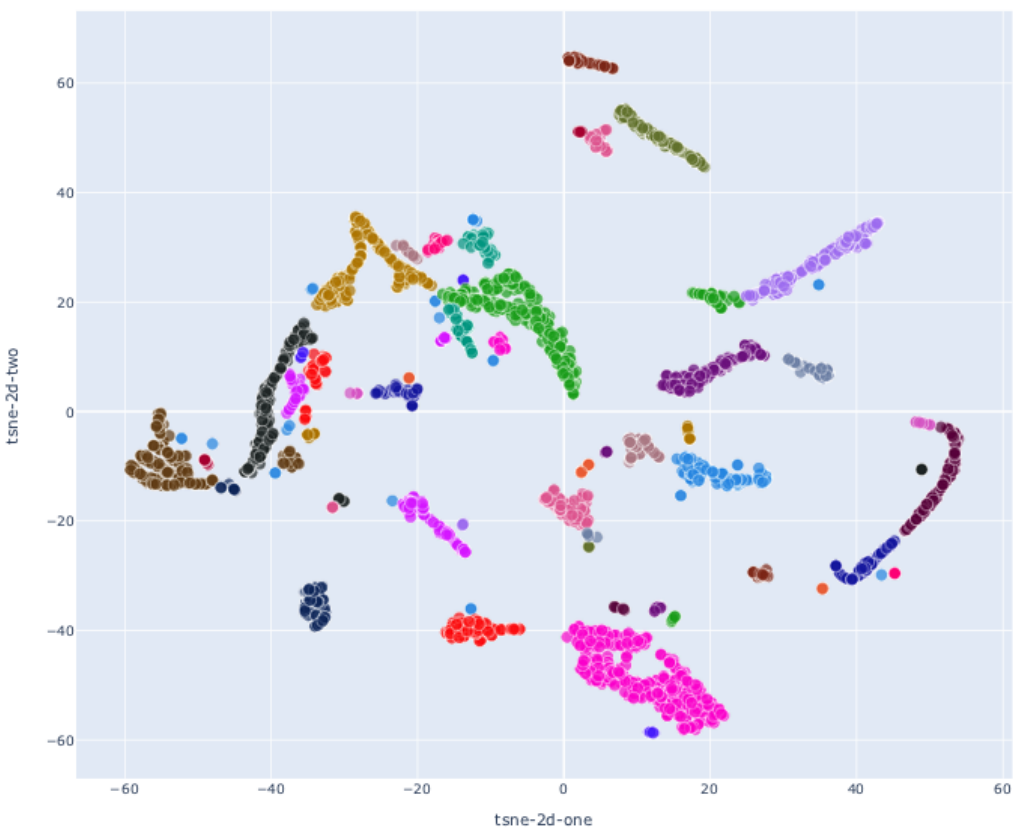

Figure S1. T-SNE of CrystalNN vectors for all 3814 two-dimensional materials in C2DB. The color encodes the different structural clusters defined by applying DBSCAN clustering.

As stated in the main text, the crystal prototype label provided by C2DB is not qualitatively univocal, and thus they are not proper for encoding the crystal structure. To define a structural label, we decided to adopt an unsupervised approach. The proposed method has three stages:

1. Generating Crystal Fingerprints (CFs) for all materials in C2DB;
2. Embedding of CFs into a 2D space;
3. Clustering the 2D embedded space into structural clusters;

To generate the CFs we used the local coordination information from all occupied sites within the unit cell. It is generated for each occupied site a 61-dimensional vector ($\mathbf{v}^{\text{site}}$) in which each dimension is a parameter characterizing the atom's coordination environment given by its coordination number $CN$ and its coordination environment $q$ (e.g. octahedral, tetrahedral, trigonal planar). $CN$ is given by the CrystalNN [1] method and $q$ by the method described by Zimmermann et al. [2]. Each dimension of $\mathbf{v}^{\text{site}}$ gives the likelihood $w_{CN=n}$ of the site having a $CN = n$, where $n \in [0, 13]$, and additionally its likelihood $q_{\text{CE}=m|\text{CN}=i}$ of having the $m$ coordination environment ($CE$), where the set of possible $m$ is given by $n$. As an example, the following $\mathbf{v}^{\text{site}}$ vector shows that for $CN = 2$, some of the possible coordination environments are $q_L$ for L-shaped geometries, $q_{water}$ for a water-like angle geometry and $q_{linear}$ for linear geometry.

$$\mathbf{v}^{\text{site}} = \begin{bmatrix} w_{\text{CN}=1}, & w_{\text{CN}=2}, & q_{\text{L}|\text{CN}=2}, & q_{\text{water}|\text{CN}=2}, & q_{\text{linear}|\text{CN}=2}, & \ldots \end{bmatrix}^{\text{T}}$$

The set $\{\mathbf{v}^{\text{i}}\}$ for the $i$ sites in the unit cell is then processed to generate a $\mathbf{v}^{\text{structure}}$, the CF, given by statistical values of each of the 61 dimensions of the following equation in all sites. We chose the minimum, maximum, mean, standard deviation, resulting in a 244-dimensional (7 statistical operations x 61 dimensions) $\mathbf{v}^{\text{structure}}$ for each of the 3814 materials in C2DB.

$$\mathbf{v}^{\text{structure}} = min(\{\mathbf{v}^{\text{i}}\})^{\text{T}} \oplus max(\{\mathbf{v}^{\text{i}}\})^{\text{T}} \oplus mean(\{\mathbf{v}^{\text{i}}\})^{\text{T}} \oplus std(\{\mathbf{v}^{\text{i}}\})$$

As the resulting CFs have a large dimensionality, we then used them as input for the T-SNE embedding technique [3]. We used a T-SNE perplexity parameter of 60, with the cosine distance as the metric. The result is a two-dimensional embedding that characterizes the structural differences and similarities of all materials in C2DB, as shown in Figure S1.

We then finally defined the structural clusters used as a categorical feature in the RF model (represented by the different colors in Figure S1) by employing the scikit-learn's implementation of the DBSCAN [5] clustering method. We used an epsilon parameter of 1.5 (which measures how close points should be to each other so they can be grouped into the same cluster) and 3 as the minimal number of points to form a cluster. This resulted in a total

of 56 different structural labels. After the applied filters described in Section "Results and discussion" in the main text, the total number of structural clusters dropped to 51.

## 1.3. Random Forest hyperparameters

We used the scikit-learn's [4] RandomForestClassifier implementation of random forest. The split criterion used was the Gini impurity. We used an ensemble of 1000 trees, with no maximum depth limit. All other parameters were set to default.

## 1.4. Compositional SHAP values

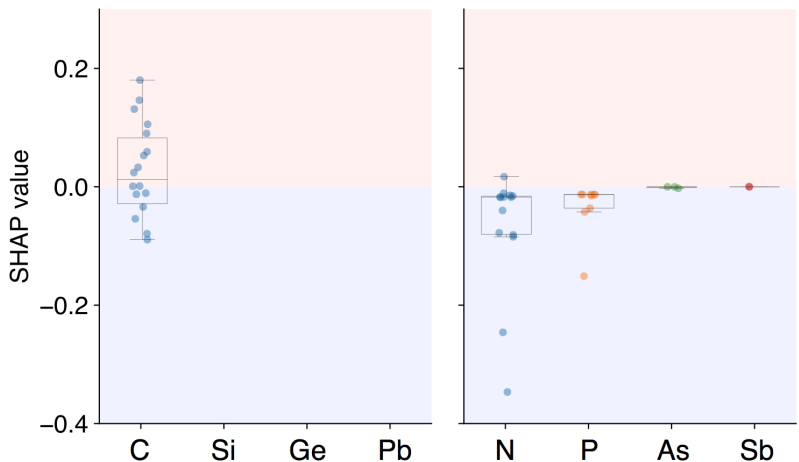

Figure 2S. Strip plots of SHAP values of compositional features for the elements from the groups of C and N. Each dot represents a calculated SHAP value for an instance/material which contains the TM in its composition. Positive (Negative) SHAP values, indicated by the red (blue) background, means that the presence of the specific element contributes to the model to classify the material as M (NM). SHAP values concentrated around zero indicates that having the element in the composition does not affect the model classification.

## 1.5. Structural clusters SHAP values

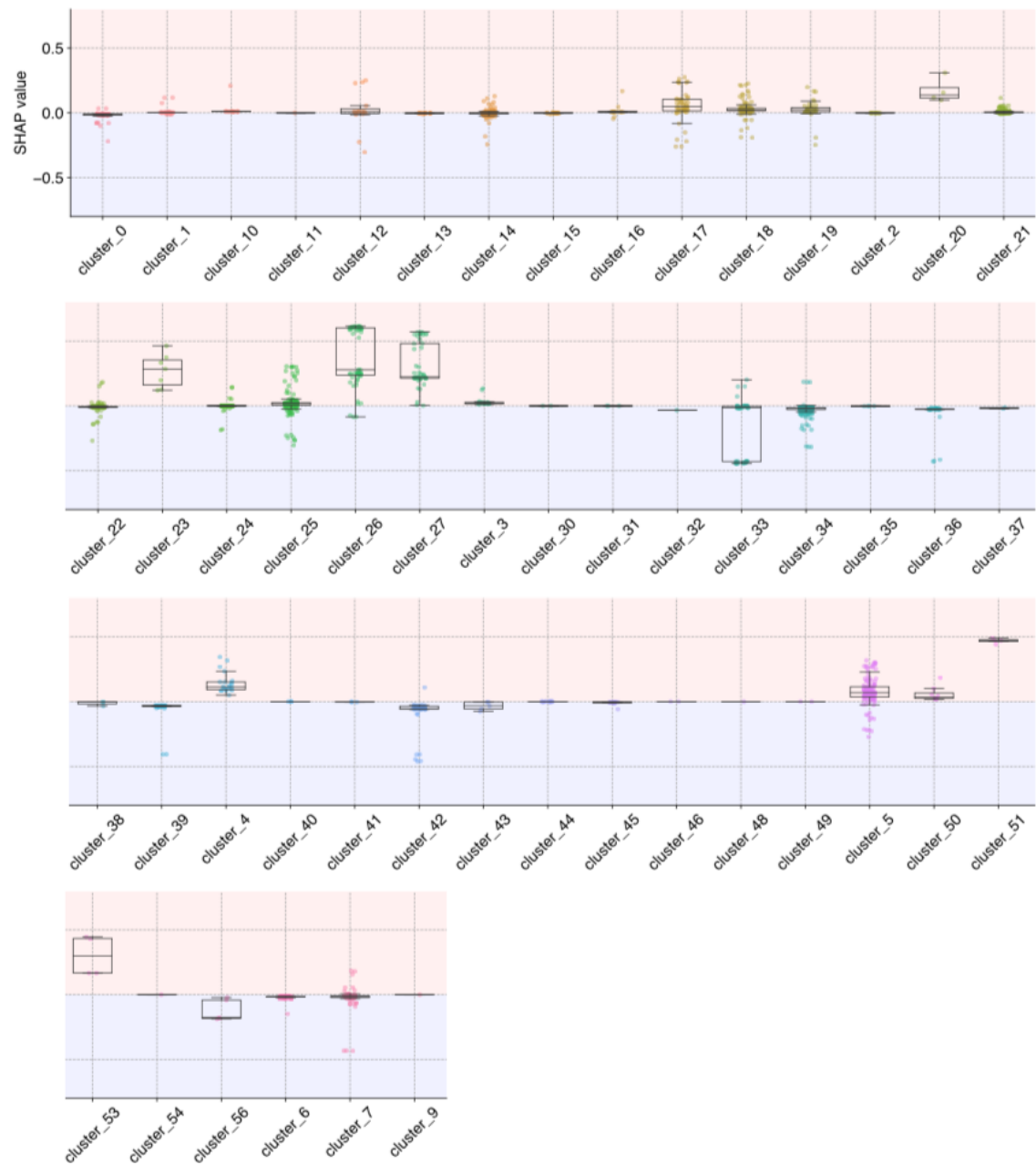

Figure 3S. Strip plots of SHAP values of structural cluster features. Each dot represents a calculated SHAP value for an instance/material which belongs to the given structural cluster. Positive (Negative) SHAP values, indicated by the red (blue) background, means that the presence of the specific element contributes to the model to classify the material as M (NM). SHAP values concentrated around zero indicates that belonging to the given structural cluster does not affect the model classification.

# 2. Machine learning classification of AFM and FM 2D compounds

## 2.1 New prototypes

The file *new_combinations.tar.gz* contains the file *new_combinations.csv,* which lists all the new prototypes, given by [structural clusters X {V, Mn, Cr, Fe, Ni, Co, Ti} X {Br, Cl, I}] non-repeated combinations.

At the root level, the POSCAR files define the crystal structures of the 169 new prototypes. It is important to disclaim that these prototypes are not optimized, i.e., they were not relaxed; they are based on a representative crystal structure from the structural clusters (randomly picked) and then ion substituted with all the 21 possible combinations of atoms.

The POSCAR files have naming formatting of *'index'_'structural cluster'_'cation'_'anion',* which are fields specified in the *new_combinations.csv* file. For example, one of the entries used for validating the AFM and FM model for $D_{3h}$ (-6m2 in the International notation) compounds is the one using Cr as cation, Cl as an anion, and structural cluster 17 as crystal structure prototype; its POSCAR file is named "*50_17_Cr_Cl*".



\end{document}